\documentclass[twocolumn, times, tight]{aastex631}

\newcommand{\code}{\texttt}
\newcommand{\modified}[1]{#1}

\usepackage{amsmath}
\usepackage{amssymb}
\usepackage{graphicx}
\usepackage{enumitem}

\SetLabelAlign{myright}{\hss\llap{$#1$}}
\newlist{where}{description}{1}
\setlist[where]{labelwidth=2cm,labelsep=1em,
                        leftmargin=!,align=myright,font=\normalfont}

\shorttitle{VarWISE}
\shortauthors{M. Paz et al.}

\graphicspath{{./}{figures/}}

\begin{document}

\title{VarWISE: Infrared Variability via NEOWISE Single Exposure Photometry}

\author[0009-0008-2229-3138]{Matthew Paz}
\affiliation{IPAC, Mail Code 100-22, California Institute of Technology, 1200 E. California Blvd., Pasadena, CA 91125, USA}

\author[0000-0003-4269-260X]{J. Davy Kirkpatrick}
\affiliation{IPAC, Mail Code 100-22, California Institute of Technology, 1200 E. California Blvd., Pasadena, CA 91125, USA}

\author[0009-0007-8625-6005]{Rajiv Uttamchandani}
\affiliation{IPAC, Mail Code 100-22, California Institute of Technology, 1200 E. California Blvd., Pasadena, CA 91125, USA}
\affiliation{Department of Physics, Loyola Marymount University, MS-8227, 1 LMU Drive, Los Angeles, CA 90045}

\author[0000-0002-3031-5279]{Troy Raen}
\affiliation{IPAC, Mail Code 100-22, California Institute of Technology, 1200 E. California Blvd., Pasadena, CA 91125, USA}

\author[0000-0002-0077-2305]{Roc M. Cutri}
\affiliation{IPAC, Mail Code 100-22, California Institute of Technology, 1200 E. California Blvd., Pasadena, CA 91125, USA}

\begin{abstract}
The Near-Earth Object Wide-field Infrared Explorer (NEOWISE) mission provides a decade of all-sky time-series data at 3.4 and 4.6 $\mu$m and an unprecedented opportunity for the discovery and characterization of variable objects. This paper presents VarWISE, a catalog of infrared-variable objects discovered within the NEOWISE single-exposure data. We employ unique methodologies, including the spatial clustering of apparitions and the adoption of novel machine learning-based variable detection (VARnet) and classification (XGBoost) to identify and characterize significant variability. The catalog includes a prediction of variable object type and best-fit period values for each object, if its variations are cyclical, along with other calculated parameters to characterize the nature of the variability. The VarWISE Pure Catalog, containing only variables of highest confidence, has $457,080$ objects, $49.81\%$ of which are new discoveries; the VarWISE Extended Catalog, containing all sources, has $1,918,082$ objects, $82.02\%$ of which are new. We discuss caveats for each variable type and highlight a few new objects found during a quick perusal of the catalogs' contents.
\end{abstract}


\section{Introduction}

Variable stars hold a special place in astronomical research. Their existence allows us to measure fundamental stellar parameters, thus providing important tests of theory. Eclipsing binaries, for example, have been used to directly measure masses and radii of stars (see \citealt{torres2010}), and variations due to starspots rotating into and out of view can, when combined with spectroscopic measurements of line broadening, give the inclination of the star relative to our line of sight if its radius is known (e.g., \citealt{vos2017}). Although such direct empirical measurements are crucial, variables have been used to unlock far bigger secrets across the cosmos as well. The realization that a Cepheid's pulsation period is a function of the star's luminosity enabled astronomers to prove that the Milky Way was not an island universe but just one galaxy among countless others (\citealt{hubble1929}). More luminous than Cepheids, type Ia supernovae have been used as standard candles to measure distances to far more distant galaxies and have revealed that the universe's expansion is accelerating (\citealt{riess1998}). 

A necessary first step in each of these processes is {\it finding} such variable sources to study further. Fortunately, we are entering a golden age of time domain astronomy wherein large areas of sky are being monitored from the ground night after night to search for objects varying in brightness or position. Current and future surveys include the Optical Gravitational Lensing Experiment (OGLE; \citealt{udalski2002}), the Zwicky Transient Facility (ZTF; \citealt{bellm2019}), the Panoramic Survey Telescope and Rapid Response System (Pan-STARRS; \citealt{chambers2016}), the All Sky Automated Survey (ASAS; \citealt{pojmanski1997}), SkyMapper (\citealt{keller2007}), the Asteroid Terrestrial-impact Last Alert System (ATLAS; \citealt{tonry2018}), the Catalina Real-Time Transient Survey (CRTS; \citealt{drake2009}), the Legacy Survey of Space and Time (LSST) from the Vera C.\ Rubin Observatory (\citealt{ivezic2019}), \modified{and the VISTA Variables in Via Lactea (VVV) survey \citep{vvv}}. Time domain surveys are also being done from space, two highly successful examples of which are Gaia (\citealt{gaia2016}) and the Transiting Exoplanet Survey Satellite (TESS; \citealt{ricker2015}).

All of these surveys, however, observe at optical or very near infrared wavelengths. In this paper, we strive to identify variable star (and galaxy) candidates using long time-baseline all-sky mid-infrared data from the space-based Near-Earth Object Wide-field Infrared Explorer (NEOWISE; \citealt{mainzer2014}), which observed at 3.4 and 4.6 $\mu$m. Such data can penetrate the dust that obscures many cloaked variable sources at shorter wavelengths -- by dust either in their own environs or dust in the Zone of Avoidance in the Milky Way. Because it is a space-based telescope, NEOWISE provides another advantage -- the photometry is very stable, particularly in W1. As shown in the NEOWISE Explanatory Supplement{\footnote{See Figure 3 at {\url {https://irsa.ipac.caltech.edu/data/WISE/docs/release/NEOWISE/expsup/sec4\_2d.html\#monitor\_zero }}}}, the variation in the W1 photometric zeropoint offsets are $<<$0.01 mag throughout the entire Reactivation mission. The decade-long time baseline of the NEOWISE Reactivation mission, late-2013 through mid-2024, provides a powerful lever arm to search for variability on hourly to multi-year timescales. \modified{Notably, \citet{kangneowise} also studied NEOWISE data from the entire mission, but instead used a coadd-derivative (\citealt{untimely}) photometric dataset and relied on unsupervised methods for variable classification.}

Briefly, the NEOWISE Reactivation mission (hereafter, NEOWISE-R) observed each patch of sky every six months. In each six-month window, a position was observed around a dozen separate times within a few-day window and was not seen again for another six months. (This is strictly true only near the ecliptic plane. Because of the ecliptic polar orbit of the spacecraft around the earth, the number of observations and the visible observing window increased toward the ecliptic poles, and the time between each observing epoch decreased.) In total, each patch of sky will have roughly twenty-one separate epochs of data in the archive\footnote{\url{https://irsa.ipac.caltech.edu/applications/Gator/}}, with typically twelve or more observations per epoch. Observations were acquired simultaneously at the W1 (3.4 $\mu$m) and W2 (4.6 $\mu$m) bands. See the NEOWISE Explanatory Supplement for more details\footnote{\url{https://irsa.ipac.caltech.edu/data/WISE/docs/release/NEOWISE/index.html}}.

\subsection{NEOWISE Data\label{subsec:neowise_data}}
Through contributed and official datasets, there are many avenues by which to access and analyze the NEOWISE mission data \citep{untimely, unwise}. However, the most granular and expansive form of the data is the NEOWISE-R Single-Exposure (L1b) Source Table, containing photometry extracted from all exposures taken by the instrument. It consists of $199,762,421,143$ rows, each row hereafter referred to as an \emph{apparition}. Each apparition contains the astrometric and photometric information pertaining to a detection of a point source on a per-exposure basis (a single point in time). Due to observational noise, the measured position of each apparition differs slightly (typically $< 1\arcsec$) from the ``true'' position of the source, necessitating methodology to collect all apparitions which pertain to a single object (see \S\ref{sec:clustering}).

Despite the lower signal-to-noise (S/N) ratio of these single-exposure data, this source table is the only way by which some variables may be effectively studied with WISE data, as other NEOWISE data products involve image co-additions or temporal smoothing which necessarily destroys some high-frequency information such as transient events and short-timescale variables. We aim to retain the maximum possible quantity of ``reliable" photometric observations of all sources.

\subsection{Data Filters}
\label{subsec:data_filters}

In order to improve data quality, we adopt some baseline filters and conventions on the single-exposure photometry prior to our first step (see \S\ref{sec:clustering}). Most importantly, we only consider apparitions with a W1- or W2-band S/N of at least $4$. This massively improves the purity of the table and removes a large portion of low-quality spurious detections, particularly of nebulosity. Additionally, we only consider apparitions without any processing pipeline-detected artifact flagging, e.g. \code{cc\_flags == 0000}. However, we find that the included artifact flag is not complete, and significant artifacts remain beyond this step. \emph{Many further, more specific filtering techniques have been employed throughout our processing of the data in an attempt to mitigate these artifacts and spurious variable detections arising thereof.} Finally, we exclude some regions \footnote{At HEALPIX partition order 5, ring partition numbers: \code{8448, 9728, 4010, 2559, 4460, 9386, 2901, 8277, 3839}} around the ecliptic poles. Due to the spacecraft's polar orbit, these regions are observed on nearly every orbit during the survey and thus have
an extremely high cadence and thus extreme data density. As these regions are relatively small in WISE, roughly $0.065 \%$ of the sky, we have made the decision to exclude them from this study, as the computational challenges are vastly different at such a high cadence. Discovery in these regions is better suited to a more narrowly focused study.

After clustering (see \S\ref{sec:clustering}) but prior to any variability analysis, we impose further filters on the data. We filter out points which originate from a frame of poor quality, generally indicative of streaked images (\code{qual\_frame > 0}), and points which have a W1-band photometry fit chi-squared value exceeding $10$, indicating a poor profile fit. After this filtering, we discard points which are greater than $4$ standard deviations away from the mean in W1, to reduce the effect of extreme outliers. Empirically, we find legitimate transients peaking above $+4\sigma$ are also observed at elevated but sub-$4\sigma$ brightness across multiple epochs, thus their detection is minimally impeded.

\section{Spatial Clustering of the Single-Exposure Source Table}
\label{sec:clustering}
The single exposure source table consists of apparitions: photometric data corresponding to a \emph{singular} timestamp. As a result, all collected data pertaining to a source on the sky are scattered as different apparitions whose association with one another as the same object is assumed via angular proximity on the sky. One option to associate these apparitions is via a crossmatch to a catalog of infrared sources such as AllWISE or unWISE using a radius between $2^{{\prime}{\prime}}$ and $5^{{\prime}{\prime}}$. However, these catalogs are not directly comparable to a single-exposure source table. For example, a very close double star may be resolved in coadded data products but not in any single exposure, or a transient event may not be visible in the coadded data at all. Therefore, we opt to rely solely on the information contained in the single-exposure source table to perform our associations.

\subsection{Density-Based Clustering}
\begin{figure}
\gridline{\fig{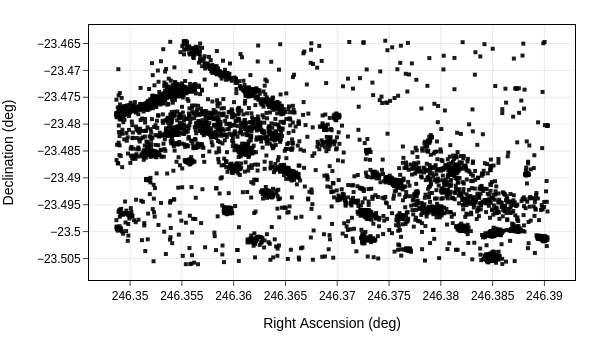}{0.4\textwidth}{(a) Original field.}}
\gridline{\fig{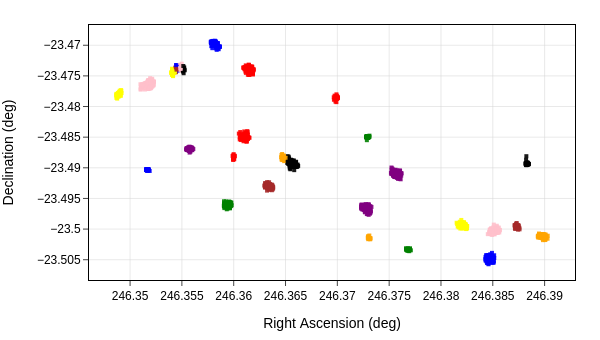}{0.4\textwidth}{(b) DBSCAN clustering.}}
\gridline{\fig{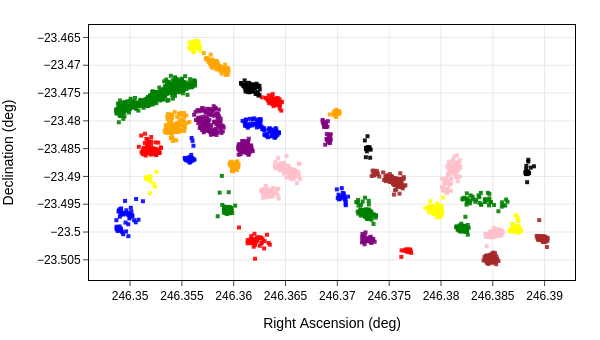}{0.4\textwidth}{(c) HDBSCAN clustering.}}
\gridline{\fig{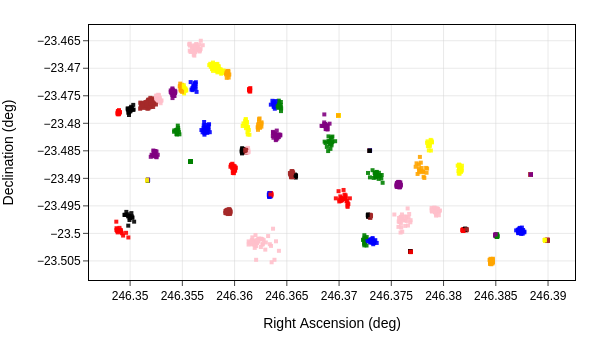}{0.4\textwidth}{(d) OPTICS clustering.}}
\caption{Comparison of clustering algorithms after tuning. Illustrated here is a small region within the $\rho$ Ophiuchi Cloud complex, containing a particularly high number of spurious apparitions due to nebulosity and artifacts from nearby bright stars. (a) Individual apparitions, shown as black points. In the following panels, the color coding shows the corresponding clusters identified by (b) DBSCAN, (c) HDBSCAN, and (d) OPTICS.}
\label{fig:clustering_comparison}
\end{figure}

There are many algorithms developed expressly for applications similar to ours, with various different advantages and drawbacks. However, after consideration of many such algorithms, we have chosen DBSCAN \citep{dbscan} for the following reasons: It does not require an initial guess for the quantity of clusters, it distinguishes points which are likely spurious versus legitimate, and it can be rigidly defined/tuned with parameters. The fact that DBSCAN operates using fixed parameters gives a tangible advantage over similar density-based clustering algorithms such as HDBSCAN \citep{hdbscan} and OPTICS \citep{optics}, as shown in Figure~\ref{fig:clustering_comparison}. Using knowledge of how the data are constructed with profile-fitting photometry allows us to make strong assumptions about which apparitions are related, and what constitutes a sufficiently dense cluster. These considerations yield an $\varepsilon$ reachability radius of $0.85$ arcseconds implemented with the haversine angular distance metric and a $\text{minPts}$ parameter of $12$ apparitions. See \citet{dbscan} or \citet{varnet} for details.

\subsection{HEALPix Partitioning and Parallelization}
At this computational scale, memory transfer overhead and capacity begin to create significant challenges. In order to manage data-in-memory and mitigate the number of distance metric evaluations (down from $N^2$), a spatial partitioning of the dataset should be adopted. We utilize the order-5 Healpix \citep{gorski2005}  partitioning of the NEOWISE Single Exposure Database that is provided by IRSA. These $12\times4^5 = 12288$ partitions cover $3.36$ square degrees apiece and range in size from $1.4$ to $108$ gigabytes. This partitioning admits a natural parallelization, heavily leveraging parquet PyArrow queries to mitigate data management overhead. However, these order-5 partitions still prove to be too large in some cases. We employ an adaptable higher-order partitioning before operating on the data, in conjunction with an edge-case routine as follows.

\begin{figure}
    \centering
    \includegraphics[width=\linewidth]{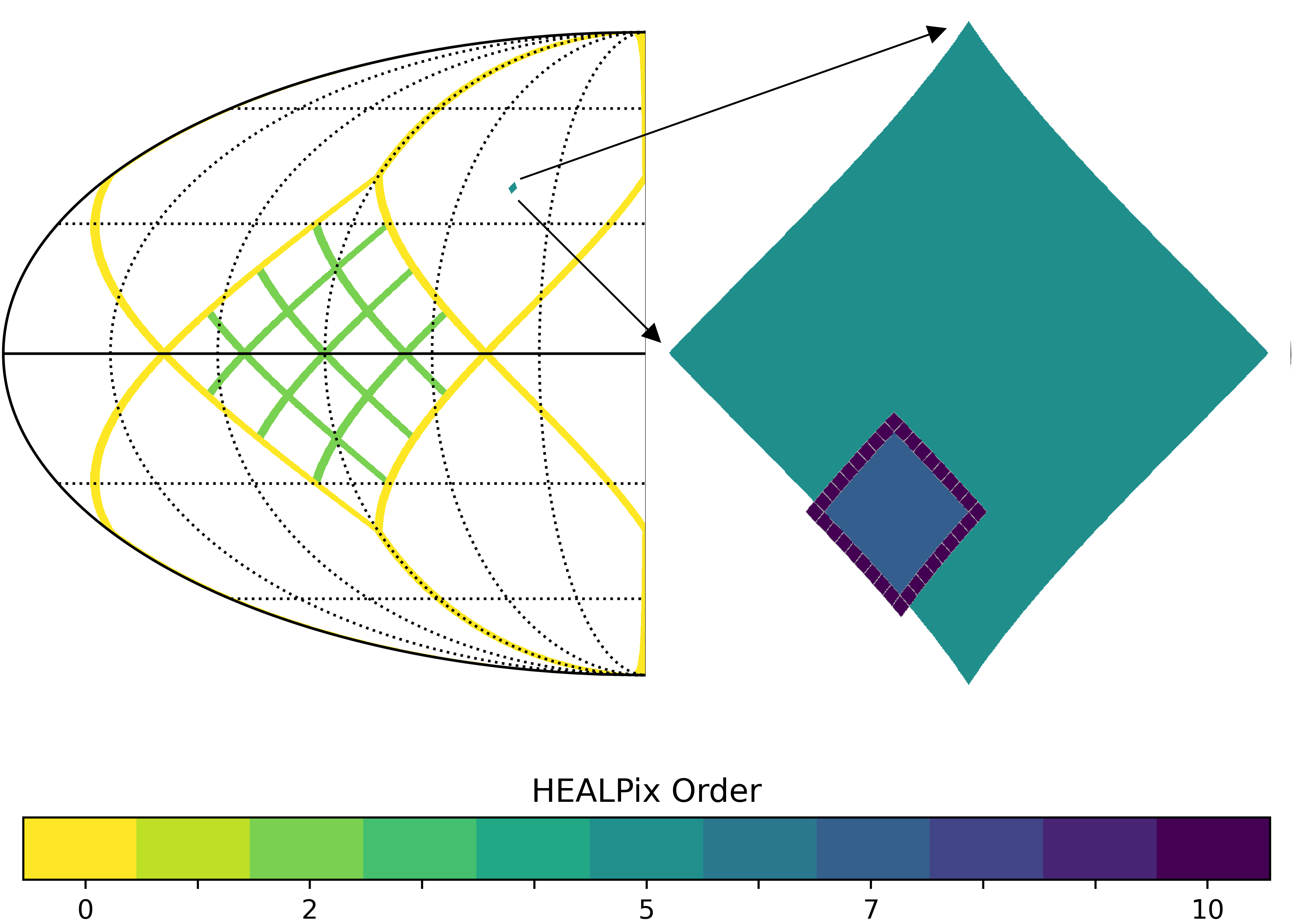}
    \caption{HEALPix partitioning. On the left is a HEALPix tessellation of the sky at order 0 (yellow) with one pixel further subdivided at order 2 (light green). A single order-5 pixel is included for scale. On the right is a zoom illustrating an order-5 pixel (dark green) representing a NEOWISE partition. A working partition is shown at order 7 (dark blue) along with its border pixels. The border pixels are shown at order 10 (purple) rather than 13 to make them more visible.}
    \label{fig:healpix}
\end{figure}

Using an order-5 HEALPix pixel as our frame of reference, we further partition into orders $6$ to $8$ depending on the density of the data as to accommodate for memory limits. In order to account for astronomical objects which appear very close to the boundaries of the working partition, we include all bordering HEALPix pixels at an order of $13$, affording roughly $25$ arcseconds of padding. This is illustrated in Figure \ref{fig:healpix}. We only record clusters whose centroids \footnote{We calculate the centroid of a cluster of apparitions using the mean of 3-dimensional unit vectors on the celestial sphere, subsequently projecting back onto right ascension and declination.} lie within the current working partition, not within the added border padding, as not to double-count any clusters.

\subsection{Results Yielded from Clustering}

In total, clustering was completed in a wall time of $91$ hours across $308$ cores (Xeon 2nd Gen), for a total of $28,028$ core-hours. The operation yielded $456,124,525$ clusters. These objects correspond to $/,458,937,050$ rows out of a total $188,876,840,852$ in the single-exposure table\footnote{This number differs from the total given in Section~\ref{subsec:neowise_data} because we used only Year 1 through Year 10 data, as the (partial) Year 11 data were not yet available at the start of this project.}. This is in contrast to the $747,634,026$ objects contained in the official AllWISE source catalog \citep{allwise_expsup}. We acquire fewer sources than this catalog in part due to source confusion -- density-based clustering will not be able to separate very proximal point clouds or point clouds that were generated from a consistently unstable photometric fitting (unresolved doubles). AllWISE is also based on detections made and extracted from all exposures covering an object, whereas VarWISE is based on the depth of the single-exposures, and this will be significantly shallower than AllWISE. Single-exposures also thus not afforded the luxury of consistent detection of objects below ${\sim}16.0$ magnitude in W1, which accounts for $422,263,350$ objects in the AllWISE catalog. Perhaps the most significant deficiency in source density from clustering arises on the Galactic plane near the Galactic Center. In this region, source confusion grows to an extreme level, to the point where the vast majority of sources \modified{are} not resolved in single exposures. Image coaddition affording higher spatial resolution in this region, coupled with adapted profile-fit photometry, allows AllWISE (and other coadded products) to resolve objects in the most dense regions of the sky. This bias in our prior distribution of sources is responsible for an artificial thinning of variable detection density near Galactic latitude $0^\circ$.

\section{Variable Detection}

\subsection{First-pass Variable Flagging}
\label{subsec:varflagging}
Included with the release of AllWISE data \citep{allwise_expsup} was a variable flag corresponding to a rather deep view of WISE-visible objects in the much shorter duration of the primary WISE survey. However, a significant portion of all NEOWISE-detectable infrared variables would be excluded from AllWISE, due to its shorter observational time span of roughly one year (07 Jan 2010 through 01 Feb 2011). Therefore, we devise our own methodology to detect variable sources.

We utilize VARnet \citep{varnet}, a deep learning and signal processing model formulated expressly for the purpose of analyzing light curves for low-SNR variable phenomena, highly applicable to the task of NEOWISE variable analysis. Specifically, we use the same VARnet model as is trained in \citet{varnet} as a first-pass classifier, categorizing each source into one of three purely morphological categories:
\begin{enumerate}
    \item Null: No apparent variability.
    \item Transient: Variability attributable mostly to a single event.
    \item Continuous Variable: Variability that continues throughout the entirety of the NEOWISE light curve.
\end{enumerate}

We discard null classifications, but separate transient and continuous variable classifications for further subclassification.

We have made the simplifying choice to only analyze W1-band variability in this step. W1 photometry is consistently deeper, and we find that variables in W2 are almost always detectable, if not more apparent, in W1.

After evaluating all $456$ million clustered objects with VARnet, $40$ million objects are flagged as either transient or continuous variables. Through inspection, we determined that the flagged objects contained legitimate variables at high confidence scores, but, as confidence decreased, also contained numerous detections that were either spurious or very low S/N. Alongside confidence thresholding, we also checked against a validation set of variable objects to derive traditional statistical thresholds that aid in the purification of this list:

\begin{equation*}
    \text{For cont. variables: }\qquad \operatorname{iqr}(J_{W1}) > (3+2\lambda)\overline{\sigma_{J_{W1}}}
\end{equation*}
\[ {\rm if} \; n_{J_{W1}} > 120\]
\begin{equation*}
    \text{For transients: }\qquad \operatorname{max}(J_{W1}) > (5+5\lambda)\overline{\sigma_{J_{W1}}}
\end{equation*}
\[ {\rm if} \; n_{J_{W1}} > 20\]
where
\begin{itemize}
    \setlength\itemsep{0.1em}
    \item $\operatorname{iqr}$ is the interquartile range.
    \item $J$ is the flux time series set.
    \item $n$ is the number of elements in the set.
    \item $\sigma_J$ is the flux error time series set.
    \item $\lambda$ is a dynamically scaled parameter.
\end{itemize}

We also observed a significant increase in the frequency of false positive detections nearing the Galactic plane. This is in part simply due to an increase in source density; however, the effects of source confusion from crowding become very apparent approaching the Galactic Center. Therefore, to be more strict, we linearly scale our threshold variable, $\lambda$, to run from 0 at Galactic latitude $\pm 10^\circ$ to 1 at Galactic latitude $0^\circ$.

After filtration using these thresholds, we acquired a purer total of $2,028,986$ variable candidates, $73,490$ of which were classified as transient events, and $1,955,496$ as continuous variables. A final filter removes sources that saturate at $W1$ for magnitudes brighter than $7.5$ mag. In \S\ref{sec:classification}, we attempt to make a classification for all $1,918,082$ objects that remain.

\section{Classification}
\label{sec:classification}

\subsection{Taxonomy}
\label{subsec:taxonomy}

Our taxonomy consists of nine classes\footnote{\modified{Along with a tenth, non-physical \code{unclear} class used as a fallback for uncertain classifications ($\code{confidence} < 0.4$).}}, spanning multiple types of phenomena. Our classes are primarily chosen based on our ability to reliably discern -- from other classes and from bogus variability -- their light-curve morphology, either from the raw light curve or a phase-folded one. Most of our classes encompass or merge more specific populations of astronomical objects, but those subclassifications would either degrade the reliability of some of our predictions, or too heavily rely on auxiliary data which may not be available for some WISE (infrared) objects. One example is using the Gaia parallax to calculate intrinsic luminosity. Table \ref{table:classes} outlines each class, a list of significant subtypes that are encompassed by the class, a short description, and citations to prior catalogs from which we have drawn training examples.

\begin{deluxetable*}{lllll}
\tablecaption{Overview of VarWISE variable types \label{table:classes}}
\tablehead{
\colhead{Label} & 
\colhead{Type} & 
\colhead{Subtypes included} & 
\colhead{Description} &
\colhead{Catalogs used as} \\
\colhead{} & 
\colhead{} & 
\colhead{within this type} & 
\colhead{} &
\colhead{training sets} \\
}
\startdata
    cep & Cepheid Variable & Cepheid type I and type II,  & Pulsating stars exhibiting & \citep{chenztf}, \\
        & & $\delta$ Sct-type & long-period, sawtoothed & \citep{gaiacepheids},  \\
        & & & variability & \citep{rimoldinigaia},  \\
        & & & & \citep{chenneowise} \\
    rr & RR Lyrae-type Pulsator & RRab, RRd & Pulsating stars exhibiting  & \citep{chenztf}, \\
        & & & short-period, sawtoothed & \citep{gaiarrlyrs}, \\
        & & & variability & \citep{rimoldinigaia}, \\
        & & & & \citep{chenneowise} \\
    lpv & Long Period Variable & Mira, semi-regular, RV Tauri- & Pulsating stars with continuous  & \citep{gaialpvs}, \\
        & & type variables & and significant variability,  & \citep{chenztf}, \\
        & & & often reddened& \citep{rimoldinigaia}  \\
    cv & Cataclysmic Variable & Novae, eruptive variables, YSO & Transient activity not originating & \nodata \\
        & & outbursts & near an extragalactic source &  \\
        & & & &  \\
    ea & Algol-type Eclipsing & Detached eclipsing binary & Transit signals, typically exceeding  & \citep{chenneowise},  \\
        & Binary & systems & 24h in period, without continuous & \citep{chenztf} \\
        & & & variability otherwise & \\
    ew & W Ursae Majoris-type  & a-type, w-type, $\beta$ Lyr-type & Transit signals, typically $<$ 2d  & \citep{chenneowise}, \\
        & Eclipsing Binary & eclipsers & in period, showing continuous & \citep{chenztf} \\
        & & & sinusoidal variability \\
    yso & Young Stellar Object & Many types of protostars, & Continuously and significantly & \citep{rimoldinigaia} \\
        & & somewhat contaminated by & varying, often reddened \\
        & & irregular variables & \\    
    agn & Active Galactic Nucleus & QSO, Seyfert, AGN, blazar & Continuous and irregular variability & \citep{gaiaextragalactic} \\
        & & & consistent with accretion. \\
    sn & Supernova & Supernovae of type I and II, tidal & Transient activity originating near  & \nodata \\
        & & disruption events & a known extragalactic source &  \\
        & & & &  \\
\enddata
\end{deluxetable*}

\subsection{Crossmatch-Determined Classes \code{cv} and \code{sn}}
\label{subsec:egal-xmatch}
Regarding the classification of variable objects as cataclysmic variables (CV) or supernovae (SN), our approach is relatively simple. Given that WISE observes very few stellar phenomena outside our own Local Group, if we can identify a transient event with a known galaxy, we can sensibly assign it the class of SN. Conversely, if we find that a transient event lies within our Local Group, it is likely to be some sort of CV-related event, e.g., stellar nova or young stellar object (YSO) eruption. 

Thus, we perform a crossmatch to known extragalactic sources using the Gaia DR3 catalogs of galaxy candidates and QSO candidates, purified using the query constraints prescribed in \citet{gaiaextragalactic}. Notably, the prescribed filters avoid the Galactic plane for purity, meaning that we acquire no \code{sn} detections there. We select a crossmatch radius of $2^{{\prime}{\prime}}$, as determined by an analysis of angular distances between known WISE extragalactic objects and the aforementioned Gaia extragalactic objects. 

Association with an object in one of these extragalactic catalogs would also be a reasonable method to identify active galactic nuclei (AGN). However, WISE is particularly well suited to detecting these objects as new discoveries, and we are able to effectively identify them using other characteristics. By not performing the association check, we maximize the likelihood of new AGN discoveries.

\subsection{Acquisition of Variable Source Training Examples}

After the step taken in \ref{subsec:egal-xmatch}, all remaining objects are continuously variable stars and galaxies. The remainder of our flagged variable objects will again be subject to supervised machine learning in order to determine which of the classes \code{cep}, \code{rr}, \code{lpv}, \code{ea}, \code{ew}, \code{yso}, or \code{agn} they most nearly belong to. Therefore, we need to acquire examples corresponding to each type in order to train a model. We have primarily chosen to use objects detected in the generalized Gaia variability catalog courtesy of \citet{rimoldinigaia} and the ZTF source classification project courtesy of \citet{chenztf}. Supplementing these catalogs, we also make use of Gaia specific object studies \citep{gaialpvs, gaiarotators, gaiacepheids, gaiarrlyrs} for classes \code{cep}, \code{rr}, and \code{lpv}, respectively. \citet{chenneowise} also provides a useful set of known periodic variables that are visible in NEOWISE data (limited, up to 2018), that we incorporate into classes \code{rr}, \code{cep}, \code{ea}, and \code{ew}. 

We perform a crossmatch between our list of NEOWISE variables and objects in these prior catalogs. A radius of $2^{{\prime}{\prime}}$ was determined as a rough compromise between the relatively large WISE pixels of $6^{{\prime}{\prime}}$ and the finer Gaia and ZTF resolutions. Even for the case where Gaia or ZTF is able to resolve multiple objects where NEOWISE cannot and where only one of those objects is truly variable, one might hope that the signal of the one variable source in the blend is still sufficiently strong in NEOWISE photometry that we detect it. However, in some cases, these prior catalogs report variable objects very close to one of our false-positive detections, leading to poor training data. Prior to training, therefore, we employ some quality-assurance statistics that, while reducing the number of examples available to us, remove abnormally poor training examples. 

\subsection{The XGBoost Algorithm}

For source classification, our algorithm of choice is XGBoost\footnote{\url{https://github.com/dmlc/}} \citep{xgboost}. We opt for a tree-based algorithm, as it is best equipped to analyze a set of features of many different types. XGBoost has been shown to be relatively easy to train and is particularly well suited to classification problems in which some classes have significantly fewer examples than others.

In short, the XGBoost algorithm works by constructing a series of regression trees, where the output of each tree in the sequence attempts to predict and account for the errors of all previous trees. Through this technique, titled gradient boosting, XGBoost is able to perform a logistic regression to the training set of data while mostly minimizing overfitting. The following subsections (\S\ref{subsec:periodsearch}, \ref{subsec:features}) outline the origin and choices of features that we provide to the XGBoost algorithm in order to evaluate the type of each object, and the cumulative list is displayed in Table~\ref{tab:xg-features}.

\begin{table*}[ht]
\centering
\caption{Features provided for XGBoost training and inference.}
\label{tab:xg-features}
\begin{tabular*}{\textwidth}{@{\extracolsep{\fill}}lll}
\hline
\textbf{Name} & \textbf{Category} & \textbf{Description} \\
\hline
\code{w2color} & Colors / Physical & W2 - W1 band color \\
\code{w3color} & Colors / Physical & W3 - W1 band color \\
\code{w4color} & Colors / Physical & W4 - W1 band color \\
\code{jcolor} & Colors / Physical & J - W1 band color \\
\code{hcolor} & Colors / Physical & H - W1 band color \\
\code{kcolor} & Colors / Physical & K - W1 band color \\
\code{period\_1} & Periodicity & Best-fit period \\
\code{period\_2} & Periodicity & Second best-fit period \\
\code{period\_1\_sig} & Periodicity & Significance of best-fit period \\
\code{period\_2\_sig} & Periodicity & Significance of second best-fit period \\
\code{mean\_chi\_2} & Photometry Statistics & Mean reduced profile-fit $\chi^2$ of measurements \\
\code{std\_chi\_2} & Photometry Statistics & Std. dev. of reduced profile-fit $\chi^2$ \\
\code{median\_uncertainty} & Uncertainty & Median flux uncertainty \\
\code{mad\_uncertainty} & Uncertainty & MAD of flux uncertainty \\
\code{iqr\_unc\_ratio} & Uncertainty & IQR divided by median uncertainty \\
\code{ivn} & Flux Statistics & Inverse von Neumann ratio \\
\code{gaussian\_chi\_2} & Flux Statistics & Deviation from Gaussian amplitude distribution \\
\code{i60r} & Flux Statistics & 60–40 percentile flux range \\
\code{i75r} & Flux Statistics & Interquartile flux range (75–25) \\
\code{i90r} & Flux Statistics & 90–10 percentile flux range \\
\code{median\_abs\_dev} & Flux Statistics & Median absolute deviation of flux \\
\code{skew} & Flux Statistics & Flux skewness \\
\code{kurt} & Flux Statistics & Flux kurtosis \\
\code{stetsonI} & Flux Statistics & Stetson I variability index \\
\code{stetsonJ} & Flux Statistics & Stetson J variability index \\
\code{stetsonK} & Flux Statistics & Stetson K variability index \\
\code{fourier\_rmse} & Lightcurve Morphology & RMSE of 4-harmonic Fourier fit \\
\code{fourier\_a1} & Lightcurve Morphology & Amplitude of first harmonic \\
\code{fourier\_a2} & Lightcurve Morphology & Amplitude of second harmonic \\
\code{fourier\_a12} & Lightcurve Morphology & $|a_1/a_2|$ amplitude ratio \\
\code{fourier\_a14} & Lightcurve Morphology & $|a_1/a_4|$ amplitude ratio \\
\code{fourier\_p21} & Lightcurve Morphology & Phase difference $p_2 - 2p_1$ \\
\hline
\end{tabular*}
\end{table*}

\subsection{Period Search\label{subsec:period_search}}
\label{subsec:periodsearch}

\begin{deluxetable}{ll}
\tablecaption{Variables used in the equations of \S\ref{subsec:period_search} and \ref{subsec:features}\label{tab:symbols}}
\tablehead{
\colhead{Expression} &
\colhead{Definition} \\
}
\startdata
    $\delta_i$ & Flux value of the $i$th point in the light curve, \\
    & ordered by corresponding timestamp \\
    $N$ & Number of observations $\delta_i$ \\
    $\sigma_i$ & Flux-uncertainty of the $i$th point in the light curve, \\
    & ordered by corresponding timestamp \\
    $t_i$ & Corresponding timestamp of the $i$th point, in order \\
    $p$ & Trial period \\
    $\mathcal{P}$ & A set of trial periods \\
    $\mathcal{S}$ & Period Significance / Power \\
    $\Phi^{-1}$ & Inverse Gaussian CDF function (probit function) \\
    $q_\delta$ & Quantile function of the the light curve magnitude \\
    & distribution \\
    $\tau$ & Window size \\
    $\mathcal{E}(p)$ & Plavchan smoothing error on period $p$  \\
\enddata
\end{deluxetable}

One of the most important pieces of information for any classification of variability is the best-fit period value. We conduct a comprehensive period search for all objects categorized in \S\ref{subsec:varflagging} as continuous variables, leveraging a version of the Plavchan periodogram \citep{plavchan}. 

Consider a phase-folded light curve $\delta$, $t \in [0,p)$ on period $p$. Generate a boxcar-smoothed counterpart $ \overline{\delta}_i$, with smoothing window size $\tau \in (0,1)$:

\begin{equation}
    \overline{\delta}_i = \frac{1}{\sum w_{t_i}}\sum\limits_j w_{t_i}(t_j) \delta_j
\end{equation}

where \[ w_{t_i}(t) = \begin{cases}
        1 & {\rm if~} 2|t-t_i| < p\tau \\
        0 & \text{otherwise}
    \end{cases} \]

Our periodogram value for this phase fold is equal to the squared error between the smoothed and original light curves:

\begin{equation}
    \mathcal{E}(p) = \sum_i \left(\delta_i - \overline{\delta}_i \right)^2
\end{equation}

The final periodogram power $\mathcal{S}(p)$ is simply the z-score of $\mathcal{E}(p)$ with respect to all $\mathcal{P}$.

We utilize a window size of $\tau = 0.1$, which strikes a balance between the fitting of sharp variability (EA-type eclipses) and smooth fluctuation (Cepheids, RR Lyraes). We find this periodogram to be effective across all types of periodic variability. For our set of trial periods, we construct a grid across several density regimes (Table~\ref{tab:period-grid}), for a total of $441,300$ periods. We range from $0.1$d as the minimum period for which we can consistently detect periodicity with this method, up to $1000$d, as we have under $4000$ days of time baseline.

\begin{deluxetable}{ccc}
\tablecaption{Piecewise period sampling grid\label{tab:period-grid}}
\tablehead{
\colhead{Range Start} &
\colhead{Range End\tablenotemark{a}} &
\colhead{Step Interval} \\
\colhead{(day)} &
\colhead{(day)} &
\colhead{(day)} \\
}
\startdata
0.1   & 5    & 0.000025 \\
5     & 15   & 0.00005  \\
15    & 30   & 0.0005   \\
30    & 100  & 0.005    \\
100   & 500  & 0.5      \\
500   & 1000 & 1        \\
\enddata
\tablenotetext{a}{These ranges are upper-exclusive. Thus, 999d is actually the maximum period.}
\end{deluxetable}

After acquiring significances for all periods in the trial set, we then apply a radius-excision operation to distinguish between different peaks in period significance, using a radius of 12.5\%. The first and second best periods are retained, along with the first significance value. 

\begin{equation}
\mathcal{S}_\mathcal{P}(p_i) = \begin{cases}
    0 & \min\limits_{k < i} \enspace \lvert p_k - p_i\rvert < \frac{p_i}{r} \\
    \mathcal{S}(p_i) & \min\limits_{k < i} \enspace \lvert p_k - p_i\rvert > \frac{p_i}{r}
\end{cases}
\end{equation}
where
\begin{itemize}
    \item $k < i \implies \mathcal{S}(p_k) < \mathcal{S}(p_i)$
    \item $r$ is chosen to be $8$ empirically
\end{itemize}

\subsection{Object Features}
\label{subsec:features}

\subsubsection{Statistical Features}
We provide the model with information about the dispersion of magnitudes across the light curve via 
\[i60r = q_\delta(0.60) - q_\delta(0.40)\]
\[iqr = q_\delta(0.75) - q_\delta(0.25)\]
\[i90r = q_\delta(0.90) - q_\delta(0.10)\]

Some other simple statistics provided to the model include the median absolute deviation of flux, the skew and kurtosis of flux, the median and standard deviation of the W1 profile-fit chi-squared values, and the median and median absolute deviation of flux uncertainty. We also provide the ratio between the interquartile range and median flux error to convey variability signal-to-noise. 

If we assume that an object with no real variability should be well-modeled by a Gaussian distribution around a mean, checking the validity of this assumption should be a useful metric in determining how variable the object is. Therefore we include the chi-squared value of the light curve flux distribution versus the corresponding Gaussian distribution with its mean and variance.

\begin{equation}
    \chi^2 = \int_0^1 \frac{\left(\Phi^{-1}(p) - P\{(\delta - \mu_\delta)^2 \leq p\}\right)^2}{\Phi^{-1}(p)} dp
\end{equation}

Historically, Stetson's variability indices \citep{stetsonI, stetsonJK} have been used to detect and characterize variability from the distribution of light curve flux. They are defined for a one-filter light curve as follows:

\[I = \sqrt{\frac{1}{N^2 - 1}} \sum\limits^N_{i=1} \frac{(\delta_i - \mu_\delta)^2}{\sigma_i^2} \]

\[J = \frac{\sum\limits^n_{k=1} w_k {\rm sgn}(P_k)\sqrt{|{P_k}}|}{\sum\limits^n_{k=1} w_k}\]

\[K = \frac{\frac{1}{N}\sum\limits^N_{i=1}|{\delta}_i|}{\sqrt{\frac{1}{N}\sum\limits^N_{i=1}{{\delta}_i}^2}}\]

where
\[P_k = \left(\frac{\delta_k - \mu_\delta}{\sigma_i}\right)^2 - 1\]
\[\code{sgn}(x) = \frac{x}{\lvert x \rvert}\]

and $w_k$ are sample weights, which we set to $1$.

\subsubsection{Folded-Morphological Features}

The specific morphology of a well-phase-folded waveform is an extremely dominant factor in classification \citep{ogle_atlas}. To extract these features, we adopt a simple Fourier-series approach inspired by \citet{chenztf}. Consider a light curve $\delta_i$ with phase-folded timestamps $t_i \in [0,1)$. We define a sine Fourier series with $k=4$ terms.

\begin{equation}
    \mathcal{F}_{\theta = a_0, \cdots, a_k, \phi_1, \cdots, \phi_k}(x) =
    a_0 + \sum\limits_{i=1}^k a_i\sin(2\pi i x_i+\phi_i)
\end{equation}
\begin{equation*}
    \theta = \arg \min_\theta \left|| F_\theta(t_i) - \delta_i \right||^2
\end{equation*}

The individual values of the fit coefficients for amplitude and phase do not necessarily have any intrinsic meaning; however, their ratios provide rich information about light curve morphology. For example, we expect the value of $a_{41} = a_4/a_1$ to increase as variability ``sharpness'' increases, indicating that the Fourier series needs higher frequency terms to converge on the proper waveform. We include the ratios $a_{41}$, $a_{21}$, and $p_{21}$ as features for supervised classification. Furthermore, the final root mean squared error of the Fourier series fit is included.

\subsubsection{Auxilliary Features}
\label{subsec:auxfeatures}
In addition to light curve morphology, physical parameters with information not contained in the light curve data can also be critical to determining a type. Although distance and an optical spectrum would be very powerful in determining the object's intrinsic physical parameters, we cannot rely on archived optical surveys to contain detections of all objects that are detectable in the infrared. One of the utilities of VarWISE is its ability to identify infrared-only variables, typically those obscured by dust. If we were to introduce a very powerful feature that does not generalize well to this population, we risk a severe degradation of prediction quality where that feature is unavailable. For this reason, we limit our auxilliary features to those that can be acquired from the infrared WISE/NEOWISE and 2MASS \citep{2mass} all-sky surveys.

By performing a crossmatch from our cluster centroids back to the AllWISE catalog \citep{allwise_expsup}, we primarily aim to acquire photometric data as features. From this crossmatch, the \code{W1mag}, \code{W2mag}, \code{W3mag}, \code{W4mag}, \code{Jmag}, \code{Hmag}, and \code{Kmag} are kept as features. In order to eliminate bias with regards to apparent magnitude, we use these to form colors instead, of the form \code{W1mag} $-$ \{band\}.

\subsection{XGBoost Results}

\begin{figure}
    \centering
    \includegraphics[width=\linewidth]{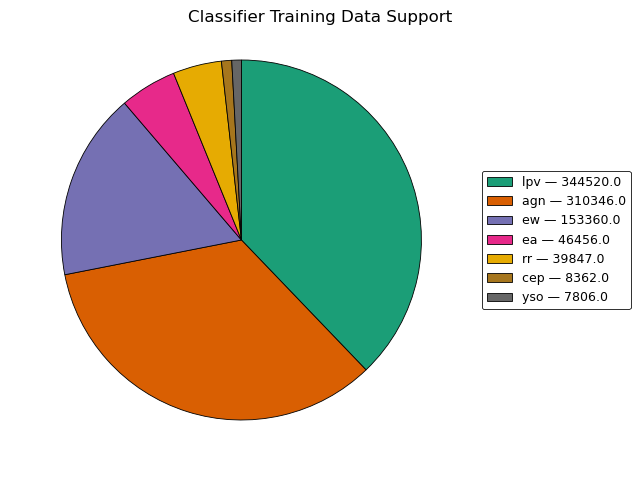}
    \caption{Training set support for the XGBoost supervised classifier. Source catalogs for these training sets are given in Table \ref{table:classes}. This distribution by type does not reflect the frequency of occurrence in NEOWISE but rather the biases in our choice of prior populations from which we are sampling.}
    \label{fig:support}
\end{figure}

\begin{figure}
    \centering
    \includegraphics[width=\linewidth]{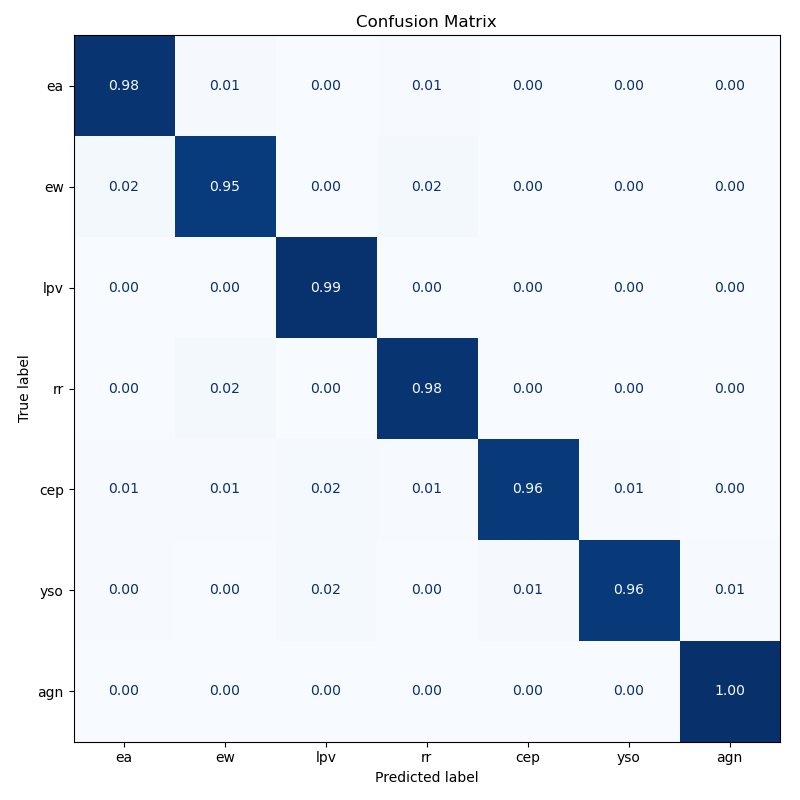}
    \caption{Performance overview of the final VarWISE XGBoost Classifier module. Confusion matrix is row-normalized.}
    \label{fig:confusion_matrix}
\end{figure}

In total, we collected $910,697$ variable objects from the literature for training (Figure \ref{fig:support}). A separate and identically-distributed \modified{validation} set of $113,840$ variables was used to evaluate the convergent performance of the model on the classification task, resulting in a macro-averaged F-1 score of $0.95$. The full confusion matrix is available in Figure \ref{fig:confusion_matrix}. These results are as satisfactory as one can hope for, but strong performance on a limited set of points does not guarantee generalization to out-of-distribution observations. One example would be rare events or phenomena that are poorly represented at optical wavelengths. The diversity and quality of prior training data remain the principal constraint for infrared-variability studies: automated classification is only as reliable as the data on which it was trained.

\section{Contents of the VarWISE Catalogs}

\subsection{Products}

VarWISE consists of two catalogs and an ancillary table, all of which are available at the NASA/IPAC Infrared Science Archive (IRSA\footnote{\url {https://irsa.ipac.caltech.edu/data/WISE/VarWISE/overview.html}, also available through
{\url {https://doi.org/10.26131/IRSA656}},
{\url {https://doi.org/10.26131/IRSA657}}, and
{\url {https://doi.org/10.26131/IRSA658}}
}.).  The larger VarWISE Extended Catalog has $1,918,082$ significant variables identified through the processes above. The column names and descriptions for this catalog can be found in Table~\ref{tab:column_descriptions}. \modified{The Gaia magnitudes for each source were taken from the VizieR version of the third data release \citep{gaiadr3mainsource}}. The smaller VarWISE Pure Catalog, described further in \S\ref{subsec:purification}, has $457,080$ objects meeting the strict criteria shown in Table~\ref{tab:purification}. The pure catalog is a subset of the extended one, and its column names and descriptions are the same as those in Table~\ref{tab:column_descriptions}.

\begin{deluxetable*}{llll}
\tabletypesize{\scriptsize}
\tablecaption{Columns in the VarWISE Catalog\label{tab:column_descriptions}}
\tablehead{
\colhead{Column Name} &
\colhead{Description} &
\colhead{Units} &
\colhead{Format} \\
}
\startdata
\code{cluster$\_$id}   & Internal VarWISE cluster identification number & \nodata & I19 \\
\code{Designation} &  VarWISE Designation                    & \nodata & A27 \\ 
\code{ra}                 &  Right Ascension,  J2000 equinox & deg       & F9.5 \\
\code{dec}              &   Declination,  J2000 equinox        & deg       & F8.4 \\
\code{vartype}             &  Predicted variable type\tablenotemark{a}& \nodata & A7 \\ 
\code{confidence}   &  Confidence in the predicted variable type\tablenotemark{b}  & \nodata & F6.4 \\
\code{variability$\_$snr} & Ratio of the span between 95th and 5th percentile of flux to the median flux error & \nodata & F6.3 \\
\code{period1} & Value of the best-fit period & day & F10.6 \\
\code{period2} & Value of the alternate fit period & day & F10.6 \\
\code{period$\_$significance} & Plavchan-type power of the phase folding for \code{period1}& \nodata & F7.3 \\
\code{suspect$\_$period} & Flag indicating whether the period is likely an alias\tablenotemark{c} & \nodata & I1 \\
\code{W1$\_$amp} & Span between the 95th and 5th percentile for the W1 magnitude & mag & F5.3 \\
\code{W2$\_$amp} & Span between the 95th and 5th percentile for the W2 magnitude & mag & F5.3 \\
\code{n$\_$obs} & Number of NEOWISE single-exposure photometric measurements at W1 & \nodata & I3 \\
\code{W1mag} & Value of w1mpro from the AllWISE Source Catalog & mag & F6.3 \\
\code{W1emag} & Value of w1sigmpro from the AllWISE Source Catalog & mag & F6.3 \\
\code{W2mag} & Value of w2mpro from the AllWISE Source Catalog & mag & F6.3 \\
\code{W2emag} & Value of w2sigmpro from the AllWISE Source Catalog & mag & F6.3 \\
\code{W3mag} & Value of w3mpro from the AllWISE Source Catalog & mag & F5.2 \\
\code{W3emag} & Value of w3sigmpro from the AllWISE Source Catalog & mag & F5.2 \\
\code{W4mag} & Value of w4mpro from the AllWISE Source Catalog & mag & F5.2 \\
\code{W4emag} & Value of w4sigmpro from the AllWISE Source Catalog & mag & F5.2 \\
\code{Jmag} & Value of j$\_$m from the 2MASS All-Sky Point Source Catalog & mag & F5.2 \\
\code{Jemag} & Value of j$\_$msigcom from the 2MASS All-Sky Point Source Catalog & mag & F5.2 \\
\code{Hmag} & Value of h$\_$m from the 2MASS All-Sky Point Source Catalog & mag & F5.2 \\
\code{Hemag} & Value of h$\_$msigcom from the 2MASS All-Sky Point Source Catalog & mag & F5.2 \\
\code{Kmag} & Value of k$\_$m from the 2MASS All-Sky Point Source Catalog & mag & F5.2 \\
\code{Kemag} & Value of k$\_$msigcom from the 2MASS All-Sky Point Source Catalog & mag & F5.2 \\
\code{Gmag} & Value of Gmag from the VizieR version of Gaia DR3 & mag & F5.2 \\
\code{Gemag} & Value of e$\_$Gmag  from the VizieR version of Gaia DR3 & mag & F5.2 \\
\code{BPmag} & Value of BPmag from the VizieR version of Gaia DR3 & mag & F5.2 \\
\code{BPemag} & Value of e$\_$BPmag  from the VizieR version of Gaia DR3 & mag & F5.2 \\
\code{RPmag} & Value of RPmag from the VizieR version of Gaia DR3 & mag & F5.2 \\
\code{RPemag} & Value of e$\_$RPmag  from the VizieR version of Gaia DR3 & mag & F5.2 \\
\code{Plx} & Gaia DR3 measured parallax & mas & F8.4 \\
\code{e$\_$Plx} & Gaia DR3 measured parallax error & mas & F6.4 \\
\code{simbad$\_$type} & SIMBAD variable source type & \nodata & A10 \\
\code{known$\_$extragalactic} & Indicates whether object is in the Gaia QSO/Galaxy catalog (Bailer-Jones  & \nodata & I1 \\
 & et al. 2023)\tablenotemark{d} &  &  \\
\code{blended$\_$source} & Likelihood of spurious variability due to WISE source confusion\tablenotemark{e} & \nodata & F4.2 \\
\code{latent$\_$artifact} & Likelihood of spurious variability due to WISE latent artifacts\tablenotemark{f} & \nodata & F4.2 \\
\enddata
\tablenotetext{a}{The predicted variable types are `cep', `rr', `lpv', `cv', `ea’, `ew’, `yso', ‘agn’, ‘sn’,  or ‘unclear’.}
\tablenotetext{b}{Runs from 0.0 (less confident) to 1.0 (highly confident). Values of 0.9 or higher are recommended.}
\tablenotetext{c}{Valid values are 0 (not likely an alias) or 1 (likely an alias).}
\tablenotetext{d}{Valid values are 0 (not in the extragalactic catalog) or 1 (in the extragalactic catalog).}
\tablenotetext{e}{Valid values are 0 (not likely a blended source) or 1 (likely a blended source).}
\tablenotetext{f}{Valid values are 0 (not likely a latent artifact) or 1 (likely a latent artifact).}
\end{deluxetable*}

A third product, the VarWISE Associations Table, provides, for every VarWISE object, a link between the \code{cluster$\_$id} value listed in the VarWISE catalogs (see Table~\ref{tab:column_descriptions}) and the individual detection \code{cntr} values from the NEOWISE-R Single Exposure (L1b) Source Table. This Associations Table allows users to identify the same cloud of data points as those used in the VarWISE analyses above, should that information prove valuable.

The summarized breakdown of WISE-observed variability from both catalogs is shown in Figure \ref{fig:varpie}. To identify sources previously published as variable objects, we conducted a query via SIMBAD\footnote{We accepted SIMBAD classes \code{EclBin LongPeriodV* LongPeriodV*\_Candidate QSO RRLyrae YSO\_Candidate Seyfert1 C* YSO Mira RSCVnV* Variable* ClassicalCep SB* Seyfert2 AGN BLLac EclBin\_Candidate AGN\_Candidate TTauri* OrionV* BYDraV* PulsV* Type2Cep CataclyV* Be* delSctV* Blazar RRLyrae\_Candidate Seyfert RVTauV* Cepheid Supernova Blazar\_Candidate Eruptive* RCrBV* Nova Supernova\_Candidate LensedImage TTauri*\_Candidate Planet\_Candidate IrregularV* Ae* GravLensSystem Variable*\_Candidate SXPheV* alf2CVnV* LensedImage\_Candidate BLLac\_Candidate Planet bCepV* Pulsar Ae*\_Candidate Cepheid\_Candidate LensingEv SB*\_Candidate gammaBurst GravLens\_Candidate RCrBV*\_Candidate gammaDorV* Mira\_Candidate EllipVar\_Candidate} to be valid confirmations of variability. See {\url{https://simbad.cds.unistra.fr/simbad/}.}}. This reveals literature for $229,365$ objects ($50.19\%$) in the VarWISE Pure Catalog and $344,720$ objects ($17.98 \%$) in the VarWISE Extended Catalog, demonstrating the huge potential for discovery in both.

\subsection{Creation of the VarWISE Pure Catalog}
\label{subsec:purification}
We include several columns related to the quality of detection and characterization so that researchers are able to perform their own down-selects from the VarWISE Extended Catalog. The columns of greatest relevance to this aim are 
\code{type}, 
\code{confidence}, 
\code{variability\_snr}, 
\code{period\_significance}, 
\code{suspect\_period}, 
\code{known\_extragalactic}, 
\code{blended\_source?}, and 
\code{latent\_artifact?}.  We have used these columns to create the VarWISE Pure Catalog that may serve a useful down-selection for general purposes. The specific criteria we employed are listed in Table \ref{tab:purification}.

\begin{deluxetable}{ll}
\tabletypesize{\scriptsize}
\tablecaption{Criteria for the VarWISE Pure Catalog\label{tab:purification}}
\tablehead{
\colhead{Description} &
\colhead{Filter} \\
}
\startdata
Not likely to be blended&  \texttt{blended\_source?} = 0\\
Not likely to be impacted by a latent& \texttt{latent\_artifact?} = 0\\
Significant signal-to-noise ratio& \texttt{variability\_snr} $> 5$\\
High confidence in assigned type or& \texttt{confidence} $> 0.8$ or\\
\ \ \ high S/N in variability& \texttt{variability\_snr} $> 10$\\
If predicted to be a periodic type\tablenotemark{a},& if \texttt{type} $\in \{\,$\texttt{ew}, \texttt{ea}, \texttt{rr}, \texttt{cep}$\}$\\
\ \ \ exhibits a good phase-fold&   $\implies$ \texttt{period\_significance} $> 12.5$ \\
Special S/N cut for the \texttt{lpv} class& if \texttt{type} = \texttt{lpv} \\
\ \ \    & $\implies$ \texttt{variability\_snr} $> 7$\\
\enddata
\tablenotetext{a}{\code{lpv} objects can be uniquely difficult or impossible to phase-fold using the WISE cadence; applying this criterion would heavily reduce the class.}
\end{deluxetable}

\section{Characterization of the VarWISE Catalogs}

\begin{figure}
\includegraphics[width=\linewidth]{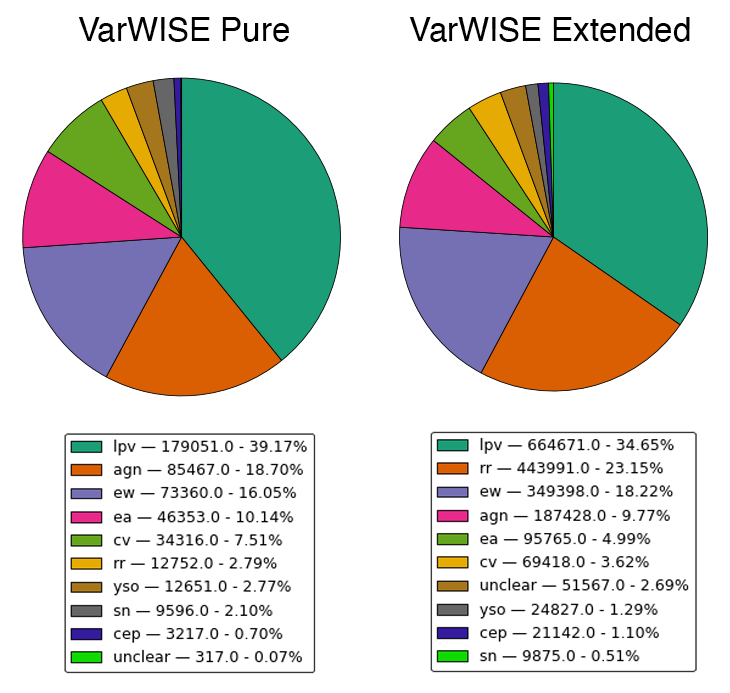}
\caption{Distribution of objects among the variable classes in the VarWISE Pure and VarWISE Extended Catalogs}
\label{fig:varpie}
\end{figure}

The distribution of variable object types in both the VarWISE Pure and Extended Catalogs is illustrated in Figure~\ref{fig:varpie}. In both catalogs, the ``lpv'' class is the most populated. In the VarWISE Pure Catalog, the second, third, and fourth most populated classes are ``agn'', ``ew'', and ``ea'', respectively, whereas in the VarWISE Extended Catalog, these are ``rr'', ``ew'', and ``agn''.

\begin{figure*}
\includegraphics[width=\linewidth]{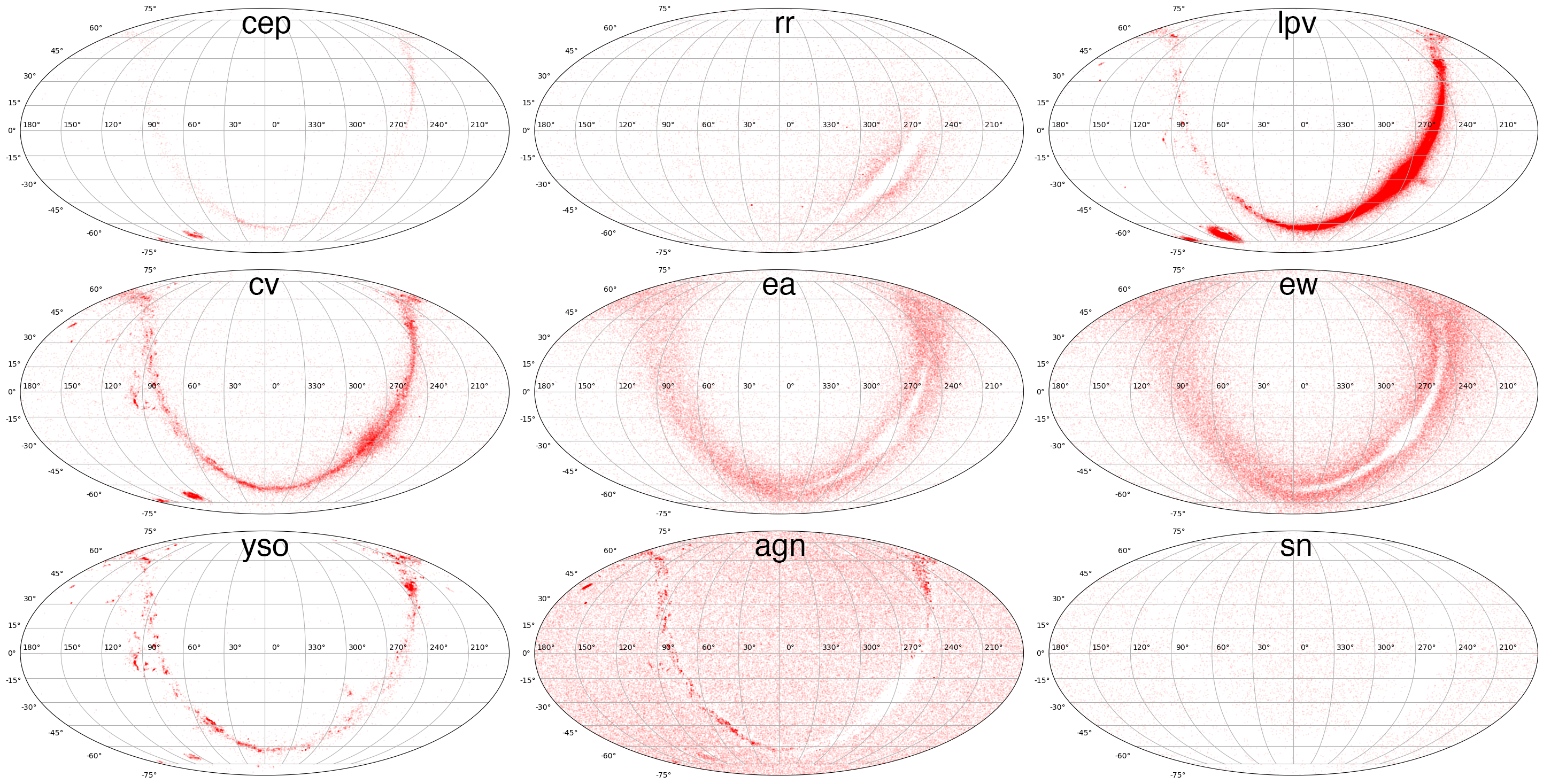}
\caption{Sky distribution of VarWISE Pure
Catalog objects among the nine variable classes. These are Mollweide projections in equatorial coordinates.
\label{fig:sky_distribution}}
\end{figure*}

The sky distribution of the nine main variable object types from the VarWISE Pure Catalog is shown in  Figure~\ref{fig:sky_distribution}. Variables associated with the Milky Way and Local Group -- ``cep'', ``rr'', ``lpv'', ``cv'', ``ea'', ``ew'', and ``yso'' -- are concentrated along the Galactic Plane and in the Magellanic Clouds, as expected. The ``cv'' and ``yso'' classes show concentrations in known star formation regions, as does ``agn'', indicating that some stellar contaminants have infiltrated the latter class; users of the catalog are encouraged to perform cuts in Galactic latitude and around the Magellanic Clouds, the Orion Nebula Complex, and the $\rho$ Ophiuchi star formation region if a purer sample of ``agn'' is required. The ``lpv'' class notably shows an overdensity in the direction of the Sagittarius Dwarf Spheroidal Galaxy. The fall-off in density toward the direction of the Galactic Center, which is particularly evident in the sky maps for types ``rr'', ``ea'', and ``ew'', is due to source confusion, as discussed in \S\ref{subsec:varflagging}. Two small regions centered at the north and south ecliptic poles, amost evident in the ``agn'' distribution, have zero density on these plots, as VarWISE does not include data in these areas (see \S\ref{subsec:data_filters}). 

\begin{figure*}
\includegraphics[width=\linewidth]{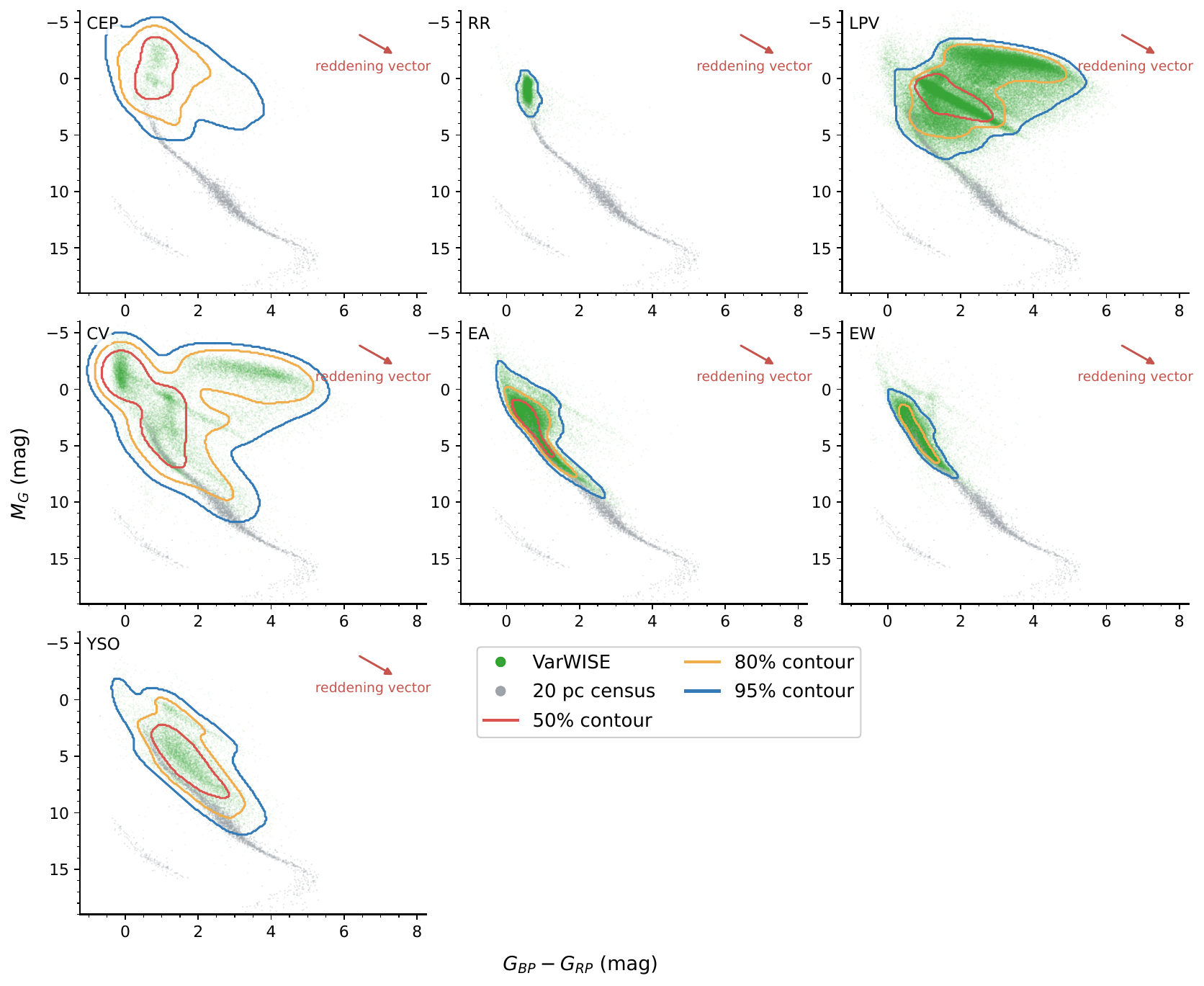}
\caption{Locations of each of our variable star classes from the VarWISE Pure Catalog in $G_{BP}-G_{RP}$ vs.\ $M_G$ color space after de-reddening corrections. Outer contours (blue) contain 95\% of the population, the next contour (orange) encloses the most concentrated 80\%, and the innermost contour (red) encloses the most concentrated 50\%. For comparison, individual objects in the 20-pc census of \cite{kirkpatrick2024} are shown in grey.}
\label{fig:G_BP-G_RP_vs_M_G_all_classes}
\end{figure*}

The locations of VarWISE Pure objects on a Gaia-based color-magnitude diagram is shown in Figure~\ref{fig:G_BP-G_RP_vs_M_G_all_classes}. These plots show only those VarWISE Pure objects with Gaia magnitudes, parallaxes, and dereddening values, so necessarily includes only those objects detected at high S/N in Gaia. Calculations for the dereddening corrections are explained further in the Appendix.

Figure~\ref{fig:G_BP-G_RP_vs_M_G_all_classes} can be compared to Figures 2 through 7 of \cite{gaia-var-paper} to see how the VarWISE classifications compare to the expected positions of known variable objects. We find that our ``cep'' class is more luminous than the ``rr'' class, as would be expected if these are true Cepheid and RR Lyrae variables. Our ``rr'' class is tightly concentrated at the absolute magnitude and color location of known RR Lyraes (Figure 3 of \citealt{gaia-var-paper}), and our ``cep'' class is much more dispersed, also in accordance with the spread of Cepheid and Type-II Cepheid types seen in Figure 3 of \cite{gaia-var-paper}. 

Objects of type ``lpv'' are concentrated in two regions of Figure~\ref{fig:G_BP-G_RP_vs_M_G_all_classes} -- one along the asymptotic giant branch, as expected, and one on a spur that represents the still improperly de-reddened extension of the red clump . (For comparison, see Figures 2 and 3 of \citealt{gaia-var-paper}.) These latter objects may not be true Long Period Variables but rather their semi-regular and irregular variable cousins.

Our ``cv'' class appears to have a mixture of objects along the asymptotic giant branch, the red giant branch (including some partially de-reddeded red clump stars), and the main sequence, so this may represent a mixture of higher luminosity Wolf-Rayet, $\gamma$ Cas, and Be-type variables (Figure 3 of \citealt{gaia-var-paper}) with lower-luminosity eruptive variables, including some YSO outbursts, nearer the main sequence. Another subgroup near the upper main sequence may, in fact, be B-type pulsators or $\beta$ Cepheids. 

Objects of type ``ea'' and ``ew'' are concentrated along the main sequence, with the ``ea'' class containing more of the cooler main sequence stars that are less frequently found in contact or near-contact scenarios (e.g., Figure 3 of \citealt{gaia-var-paper}). We have specifically not included a VarWISE class to include rotational variability, as the driver of such variability is generally inhomogeneously distributed starspots that might not show significant variations at NEOWISE wavelengths. However, other rotational effects occasionally are seen, such as ellipsoidal variations which manifect themselves as sinusoidal period signals similar to the light curves of W UMa-type eclipsers. Thus, we'd expect some contamination of our ``ew'' class by these objects. However, as both ellipsoidal variables and EWs share the same $G_{BP}-G_{RP}$ vs.\ ${M_G}$ phase space (Figures 4 and 5 of \citealt{gaia-var-paper}), this contamination is not obvious in Figure~\ref{fig:G_BP-G_RP_vs_M_G_all_classes}.

Finally, objects of type ``yso'' are found along and above the main sequence, as expected from their continuing evolution along Hayashi tracks. The location of the VarWISE ``yso'' objects appears very similar, as expected, to the location of various T Tauri classes in Figure 6 of \cite{gaia-var-paper}.

\begin{figure*}
\includegraphics[width=\linewidth]{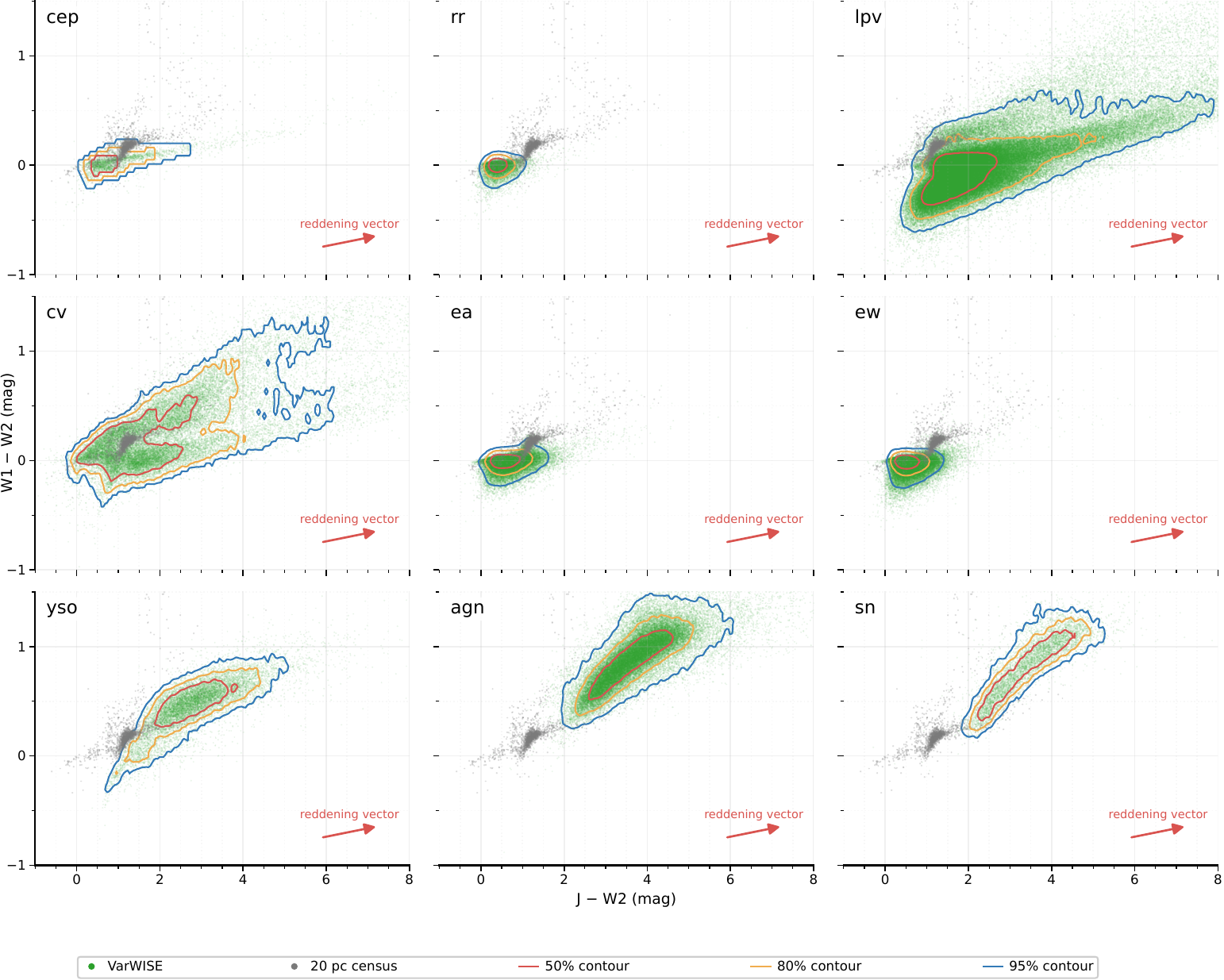}
\caption{Locations of each of our variable star classes from the VarWISE Pure Catalog in $J-$W2 vs.\ W1$-$W2 color space. Outer contours contain 95\% of the population, the next contour encloses the most concentrated 80\%, and the innermost contour encloses the most concentrated 50\%. For comparison, individual objects in the 20-pc census of \cite{kirkpatrick2024} are shown in grey.}
\label{fig:JW2_vs_W1W2_all_classes}
\end{figure*}

The $J-$W2 vs.\ W1$-$W2 diagram of Figure~\ref{fig:JW2_vs_W1W2_all_classes} shows the raw (un-dereddened) colors for objects in the VarWISE Pure Catalog. The classes showing the most reddening are ``cv'', ``lpv'', and ``cep'', which is a consequence of the fact that these objects can be seen to much higher distances -- and through larger columns of Galactic dust -- due to their brighter intrinsic luminosities, as compared to other classes such as ``rr'', ``ea'', and ``ew'', Objects in the``yso'' class also show various amounts of reddening, with the intrinsically fainter objects in the class (redder in intrinsic color) showing less reddening because their counterparts at higher extinction values will be undetected at $J$-band and thus not plotted here. Objects in class ``agn'' and ``sn'' show a combination of Galactic extinction and their own intrinsic reddening.

In the subsections below, we describe the per-class contents of the VarWISE Catalogs, along with any special per-class caveats for which users should be aware. We have made selections of catalog contents that are intended to provide a random, all-sky sampling of roughly 100 to 200 objects per class. Table~\ref{tab:analysis_column_selections} gives the selections used on the VarWISE Pure Catalog to create these lists. Examples of interesting new discoveries from the VarWISE Extended Catalog are also noted in the subsections that follow. 

\begin{deluxetable}{lcccc}
\tabletypesize{\scriptsize}
\tablecaption{Selections Used on the VarWISE Pure Catalog Columns to Create the Lists in \S~\ref{sec:analysis_ea} through ~\ref{sec:analysis_sn}\label{tab:analysis_column_selections}}
\tablehead{
\colhead{\code{type}} &
\colhead{\code{simbad\_type}} &
\colhead{\code{period1}} &
\colhead{\code{Designation}} &
\colhead{\# of}\\
\colhead{is} &
\colhead{is} &
\colhead{contains} &
\colhead{contains} &
\colhead{objects}\\
\colhead{} &
\colhead{} &
\colhead{the string} &
\colhead{the string} &
\colhead{selected}
}
\startdata
\code{ea}  & null & ``27''  & \nodata & 262 \\
\code{ew}  & null & ``39''  & \nodata & 110 \\
\code{rr}  & null & ``7''   & \nodata & 117 \\
\code{cep} & null & ``8''   & \nodata & 229 \\
\code{lpv} & null & \nodata & ``0.66''& 121 \\
\code{cv}  & null & \nodata & ``483'' & 149 \\
\code{yso} & null & \nodata & ``.09'' & 81  \\
\code{agn} & null & \nodata & ``4.58''& 84  \\
\code{sn}  & null & \nodata & ``13.'' & 212 \\
\enddata
\end{deluxetable}

\subsection{Class ``ea''\label{sec:analysis_ea}}

Of the 262 ``ea'' objects selected via the criteria in Table~\ref{tab:analysis_column_selections}, 100\% are eclipsing variables, although 10\% (27/262) have classifications that perhaps edge closer to type EW than to type EA. There are no poorly constrained periods: 92\% (241/262) have periods equal to or very close to either \code{period1} or \code{period2}. These generally differ by a factor of two; as there is ambiguity between primary-only eclipsers and dual-eclipse systems with equal-temperature components, it is not always clear which of the two periods is the correct one. The other 8\% (21/262) have periods that are a multiple of either \code{period1} or \code{period2}, and the true answers are usually two-thirds of one of these periods. No special caveats are needed for this class.

\begin{figure}
\includegraphics[width=\linewidth]{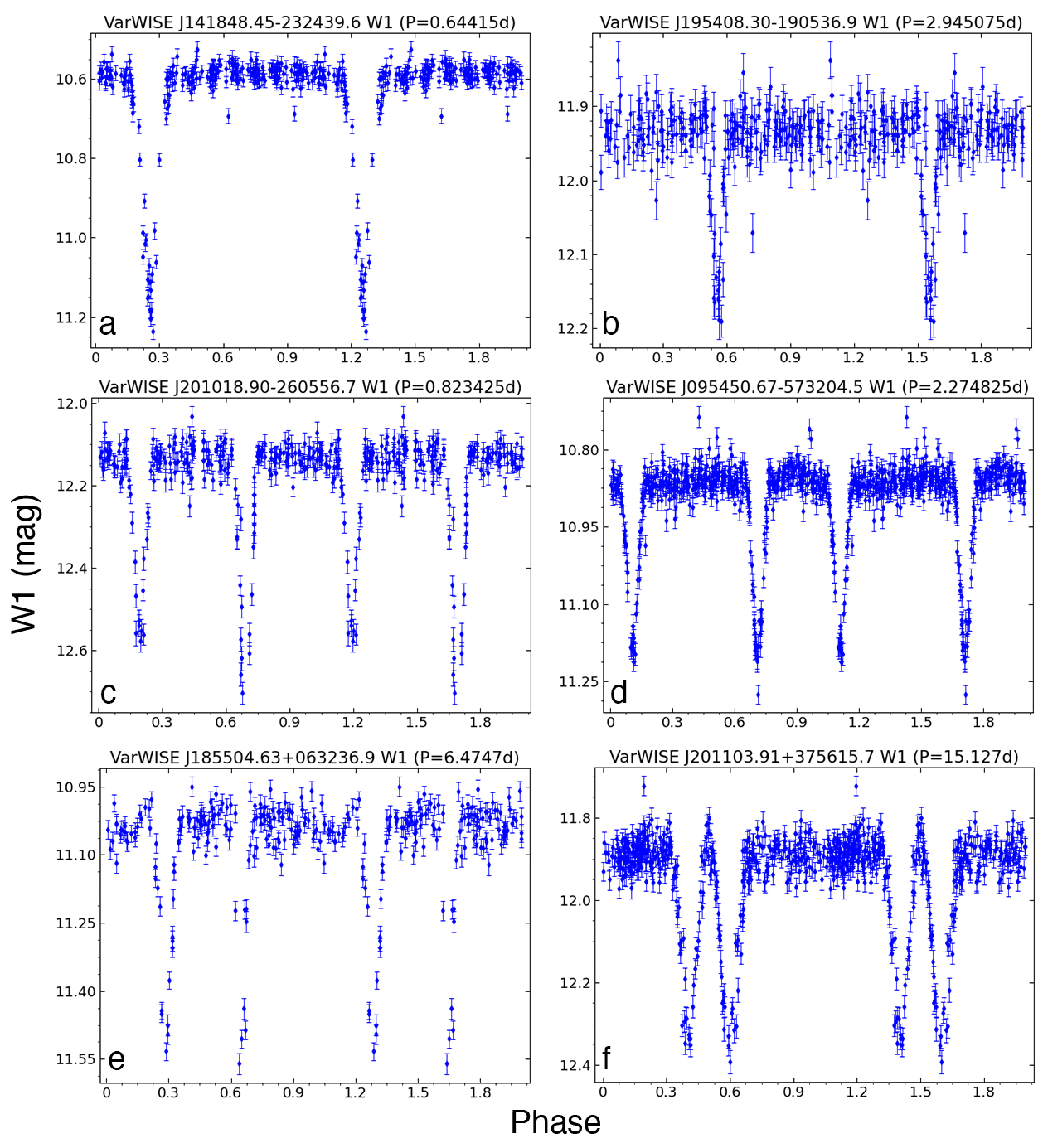}
\caption{Phase-folded light curves for six candidate ``ea'' discoveries. See text for details.}
\label{fig:discoveries_ea}
\end{figure}

Six candidate ``ea'' discoveries from VarWISE are shown in Figure~\ref{fig:discoveries_ea}. Two of these objects, VarWISE J141848.45-232439.6 (Figure~\ref{fig:discoveries_ea}a) and VarWISE J195408.30-190536.9, (Figure~\ref{fig:discoveries_ea}b), show only a primary eclipse, whereas 
VarWISE J201018.90-260556.7 (Figure~\ref{fig:discoveries_ea}c), shows clear primary and secondary eclipses. The other three objects also show clear primary and secondary eclipses, but the different time spacing between the two eclipses indicates an eccentric orbit: VarWISE J095450.67-573204.5 (Figure~\ref{fig:discoveries_ea}d) exhibits a moderately eccentric orbit, whereas VarWISE J185504.63+063236.9 (Figure~\ref{fig:discoveries_ea}e) and VarWISE J201103.91+375615.7 (Figure~\ref{fig:discoveries_ea}f) show more highly eccentric ones.

\subsection{Class ``ew''}

Of the 110 ``ew'' objects selected via the criteria in Table~\ref{tab:analysis_column_selections}, 97\% (107/110) are likely EW eclipsing variables, although there may be some small contamination by a couple of RR Lyrae variables and one object that might be more accurately classified as an EA. The three objects whose light curves do not obviously appear to be EW variables are the only three whose periods are not well constrained; all others have true periods equal to or very close to either \code{period1} or \code{period2} (70\%, or 77/110), or periods that are multiples of either \code{period1} or \code{period2} (27\%, or 30/110; at two-thirds or one-half of one of these periods).

Taking a closer look at the three oddball objects in this class, we find that all of them -- VarWISE J105605.30$-$610057.2, VarWISE J161333.82$-$345053.8, and VarWISE J190632.12$-$092814.5 -- have light curves with variability at two different timescales. A short-timescale variation is seen in NEOWISE intra-epochal periods, and a longer-term variation is seen over months to years. This multi-mode variability complicates the classification.

\begin{figure}
\includegraphics[width=\linewidth]{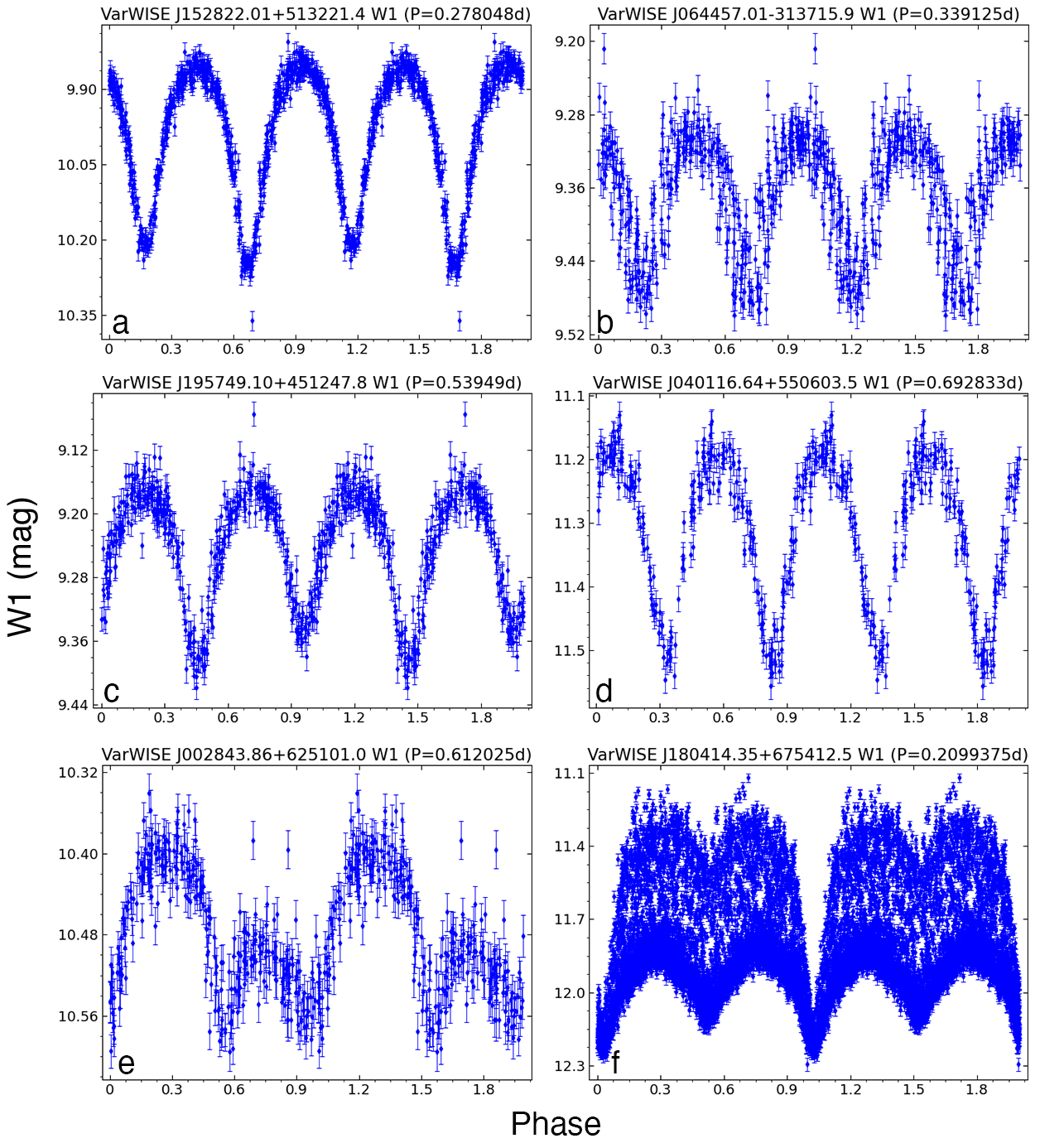}
\caption{Phase-folded light curves for six candidate ``ew'' discoveries. See text for details.}
\label{fig:discoveries_ew}
\end{figure}

\begin{figure}
\includegraphics[width=\linewidth]{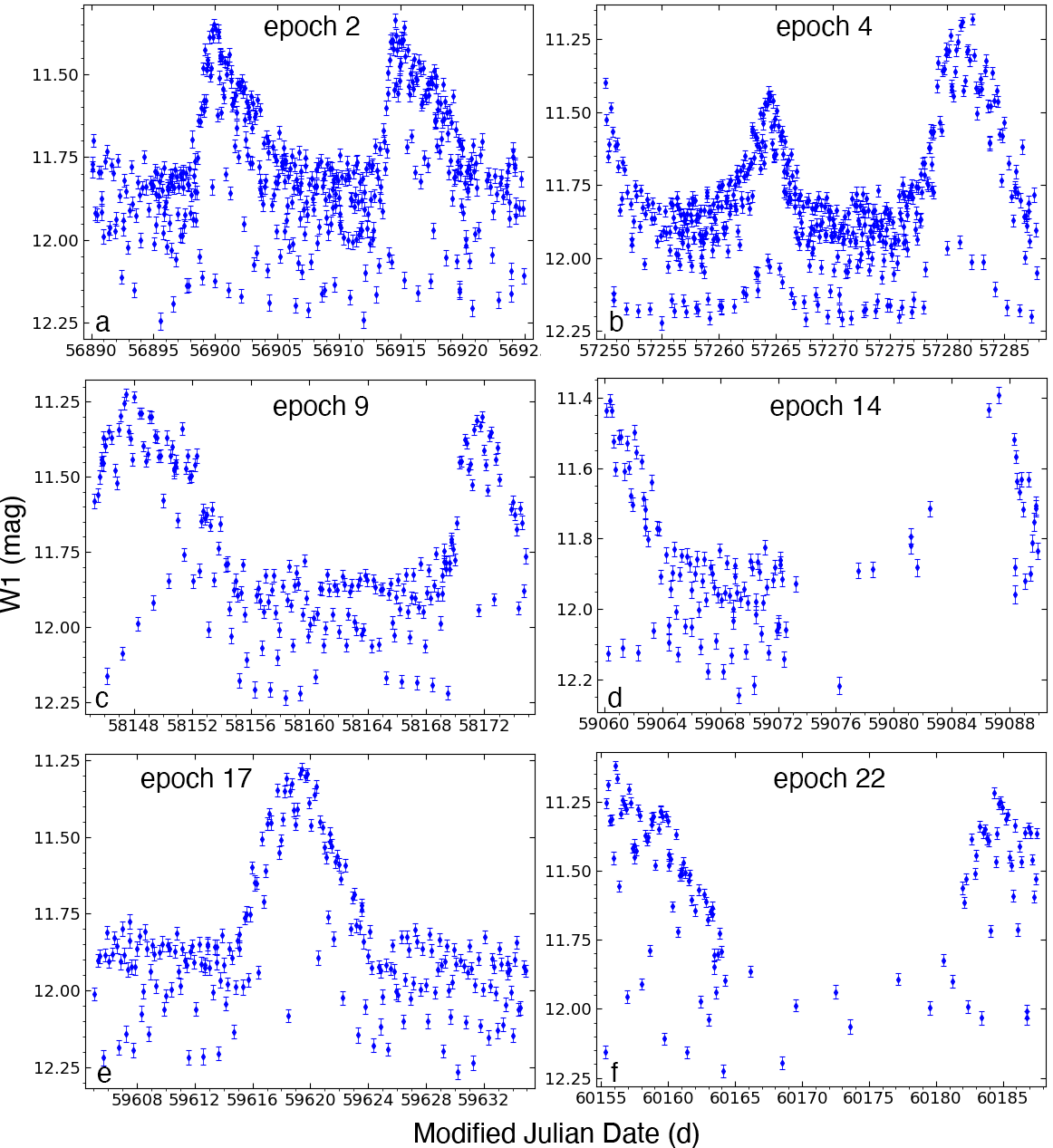}
\caption{Raw light curves of VarWISE J180414.35+675412.5 (from Figure~\ref{fig:discoveries_ew}f) at six different NEOWISE epochs.}
\label{fig:1804p6754_raw_lc}
\end{figure}

Six candidate ``ew'' discoveries from VarWISE are shown in Figure~\ref{fig:discoveries_ew}. The first four objects -- 
VarWISE J152822.01+513221.4 (Figure~\ref{fig:discoveries_ew}a),
VarWISE J064457.01-313715.9 (Figure~\ref{fig:discoveries_ew}b),
VarWISE J195749.10+451247.8 (Figure~\ref{fig:discoveries_ew}c), and
VarWISE J040116.64+550603.5 (Figure~\ref{fig:discoveries_ew}d) --
show ``ew'' variables with periods from 0.2 to 0.7d. 
VarWISE J002843.86+625101.0 (Figure~\ref{fig:discoveries_ew}e) shows an eclipsing binary exhibiting the so-called O'Connell effect, for which there is a clear difference in the heights of the maxima between eclipses (\citealt{milone1969}); the underlying causes for this light curve morphology are not yet well understood (e.g., \citealt{flores-cabrera2025, wilsey2009}).
VarWISE J180414.35+675412.5 (Figure~\ref{fig:discoveries_ew}f) shows  an eclipsing binary with a period of 0.2099375d that also shows another periodicity whose morphology changes from cycle to cycle and varies on a 14- to 18-d timescale (Figure~\ref{fig:1804p6754_raw_lc}).

\subsection{Class ``rr''}

Of the 117 ``rr'' objects selected via the criteria in Table~\ref{tab:analysis_column_selections}, 98\% (115/117) are periodic variables, most of which (83\%, or 96/115) have a true period at or very near the \code{period1} or \code{period2} value listed in the Catalog; the others (17\%, or 19/115) have a true period at a multiple of the \code{period1} or \code{period2} value. Of these 115 periodic variables, a small fraction may be W UMa-type eclipsers rather than RR Lyraes, although the morphology of the light curve alone does not make this clear. The two objects that may not be true periodic variables are WISE sources located in the Galactic Plane and have multiple Gaia detections within the WISE beam.

\begin{figure}
\includegraphics[width=\linewidth]{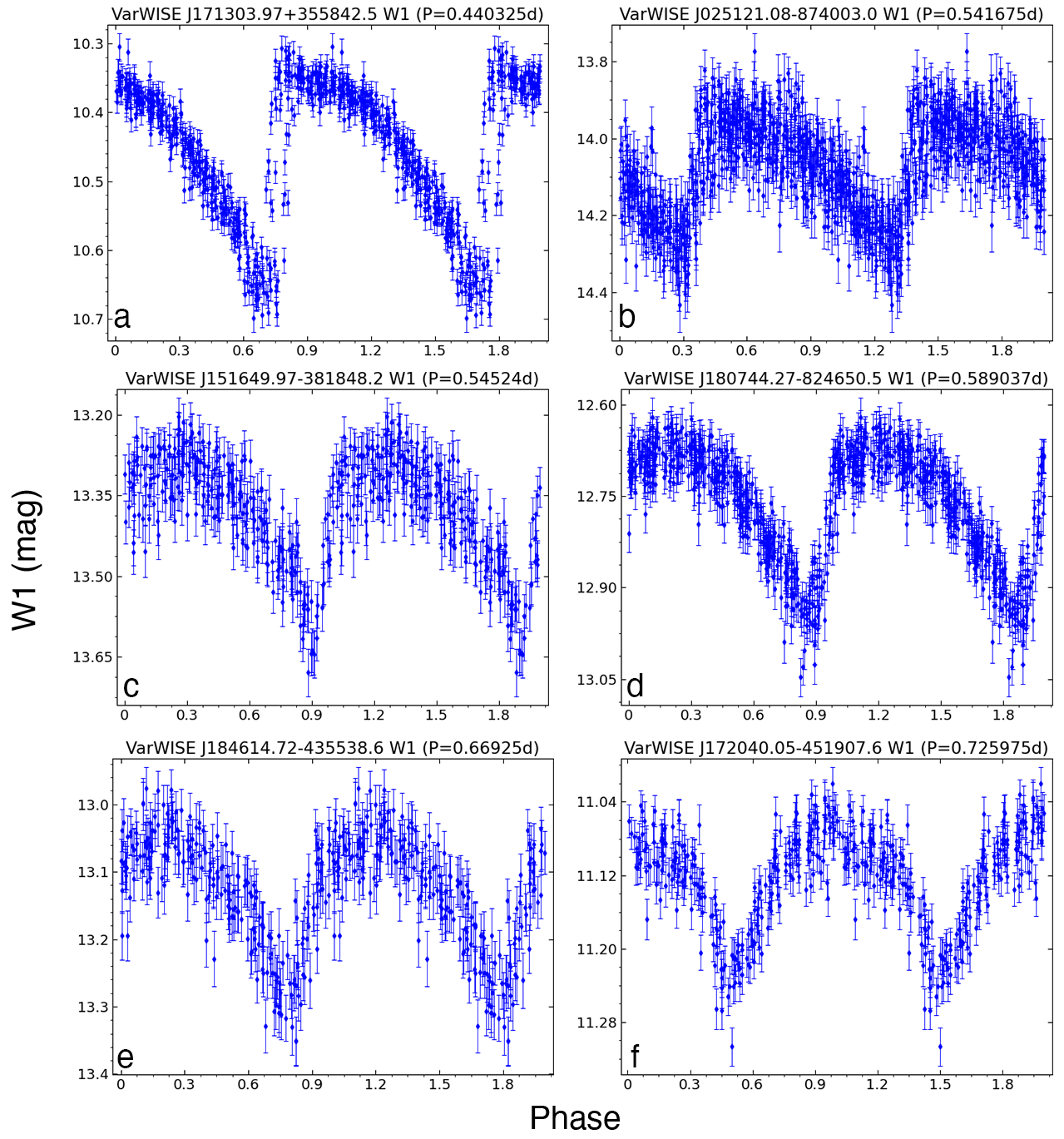}
\caption{Phase-folded light curves for six candidate ``rr'' discoveries. See text for details.}
\label{fig:discoveries_rr}
\end{figure}

Six candidate ``rr'' discoveries from VarWISE are shown in Figure~\ref{fig:discoveries_rr}. Light curves are shown in order of increasing period for objects VarWISE J171303.97+355842.5, VarWISE J025121.08$-$874003.0, VarWISE J151649.97$-$381848.2, VarWISE J180744.27$-$824650.5,  VarWISE J184614.72 $-$435538.6, and VarWISE J172040.05-451907.6 in Figure~\ref{fig:discoveries_rr}a-f, respectively.

\subsection{Class ``cep''}

Of the 229 objects selected via the criteria in Table~\ref{tab:analysis_column_selections}, 82\% (187/229) are periodic variables with a true period at or very near either \code{period1} or \code{period2} value listed in the Catalog. The others (18\%, or 42/229) have periods close to six months or a year, which is suspiciously aligned with the WISE six-month observing cadence. If these periods are real, these are more likely to be Cepheids of period $>$30 days, which are hard to distinguish in the WISE cadence, or LPVs. For others, the best periods may simply be an artifact of the observing cadence, in which case these may be CVs instead. Users are encouraged to examine the light curves and other ancillary data -- colors and, if available, absolute magnitudes -- to make a final source-by-source determination.

\begin{figure}
\includegraphics[width=\linewidth]{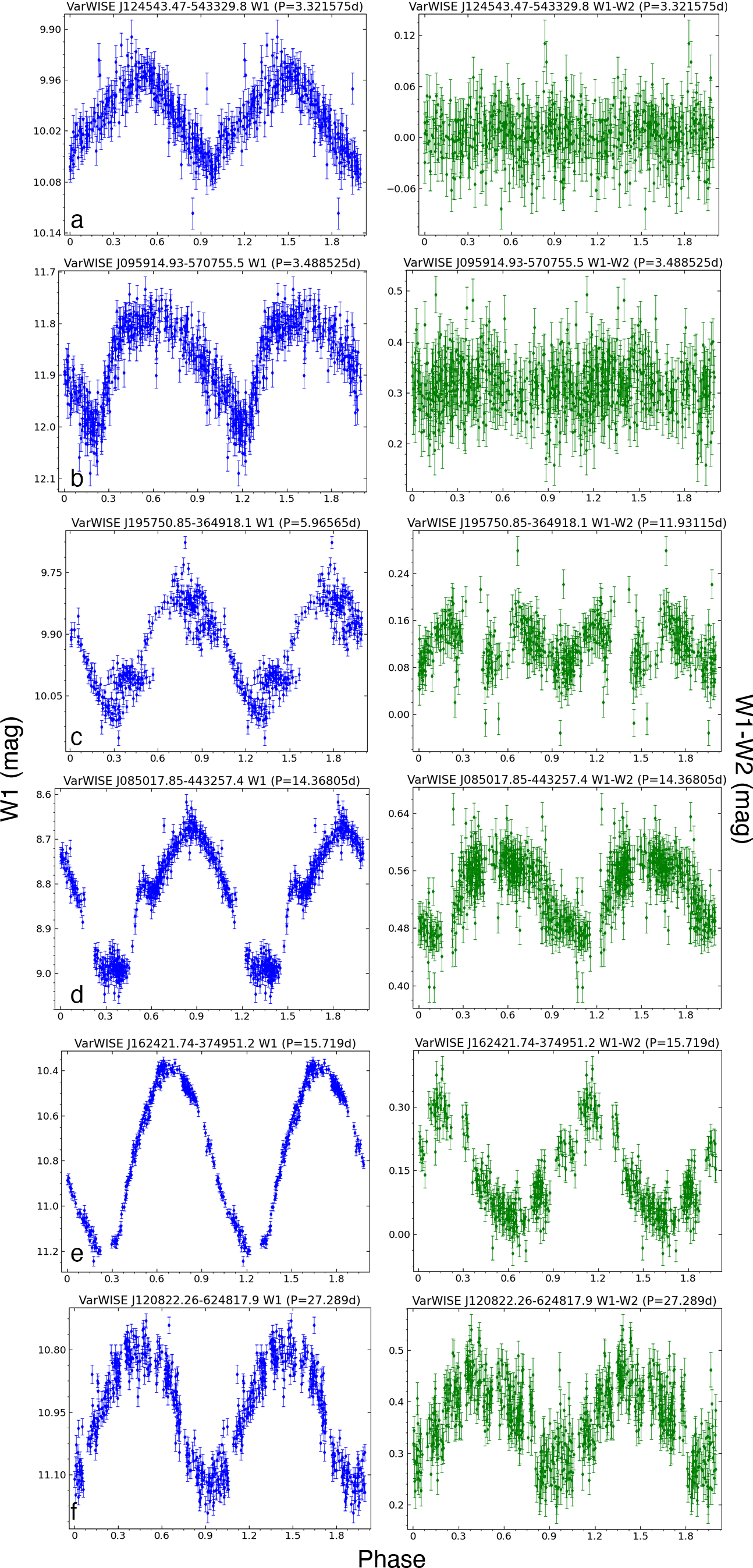}
\caption{Phase-folded W1 light curves (left column, blue) and phase-folded W1$-$W2 color trends (right column, green) for six candidate ``cep'' discoveries. See text for details.
\label{fig:discoveries_cep}}
\end{figure}

True periodic variables found in this class contain Cepheids with a variety of different light curve shapes. Figure~\ref{fig:discoveries_cep} shows six candidate ``cep'' discoveries from VarWISE, ordered by period to show a Hertzsprung-like progression of light curve morphologies (\citealt{hertzsprung1926}). Shown are VarWISE J124543.47$-$543329.8, VarWISE J095914.93$-$570755.5, VarWISE J195750.85$-$364918.1, VarWISE J085017.85$-$443257.4, VarWISE J162421.74 $-$374951.2, and VarWISE J120822.26$-$624817.9 in Figure~\ref{fig:discoveries_cep}a-f, respectively. Because NEOWISE acquired W1 and W2 imagery simultaneously, such light curves can be studied for their color variations as well, as also demonstrated in Figure~\ref{fig:discoveries_cep}.

\subsection{Class ``lpv''}

Of the 121 objects selected via the criteria in Table~\ref{tab:analysis_column_selections}, 24 (20\%) appear to be truly periodic variables, although three of these may more likely be eclipsing binaries than long-period variables. Another 94 (78\%) appear to be non-periodic, some of which are likely semi-regular or irregular variables. The rest, 3 (or 2\%), are false variables created by the nebular edge problem described in \S~\ref{sec:analysis_yso}.

\begin{figure}
\includegraphics[width=\linewidth]{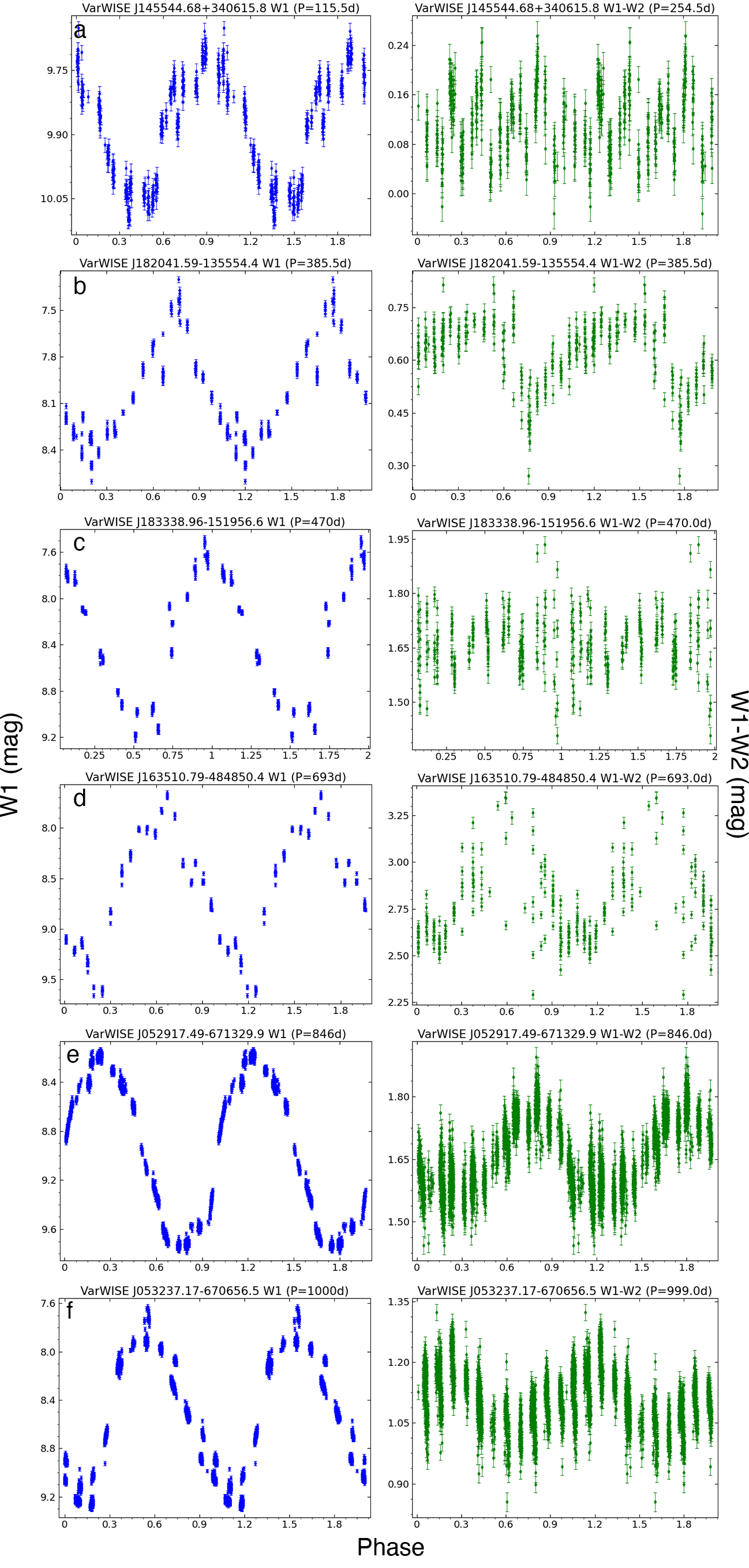}
\caption{Phase-folded W1 light curves (left column, blue) and phase-folded W1$-$W2 color trends (right column, green) for six candidate ``lpv'' discoveries. See text for details.
\label{fig:discoveries_lpv}}
\end{figure}

For illustration, Figure~\ref{fig:discoveries_lpv} shows six candidate ``lpv'' discoveries from VarWISE, ordered by increasing period. These are 
VarWISE J145544.68+340615.8, VarWISE J182041.59$-$135554.4, VarWISE J183338.96$-$151956.6, VarWISE J163510.79$-$484850.4, VarWISE J052917.49 $-$671329.9, and VarWISE J053237.17$-$670656.5 in Figure~\ref{fig:discoveries_lpv}a-f, respectively. Both the phase-folded W1 light curves and phase-folded W1$-$W2 color trends are shown.

\subsection{Class ``cv''}

Of the 149 ``cv'' objects selected via the criteria in Table~\ref{tab:analysis_column_selections}, 92\% (137/149) appear to be variable candidates of the explosive or outburst YSO varieties. A small number, 4\% (6/149), appear to be either AGN or SN given that archival PanSTARRS imaging shows a galaxy at these locations; these were not flagged as extragalactic because they do not appear in the Gaia extragalactic list. The final 4\% (6/149) appear to be spurious variables caused by either source confusion, unflagged latents or diffraction spikes, or the nebular edge problem discussed further in \S~\ref{sec:analysis_yso}.

\begin{figure}
\includegraphics[width=\linewidth]{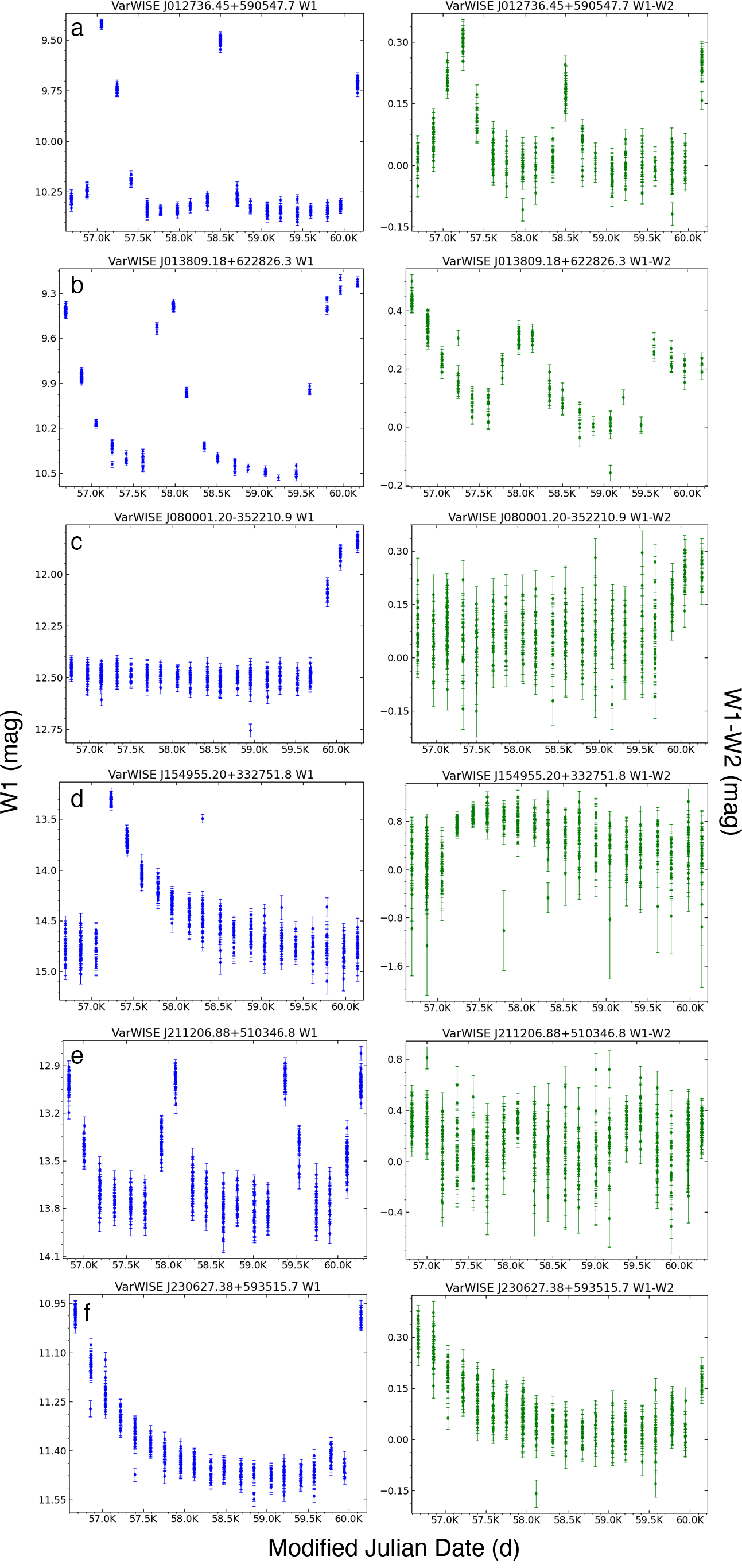}
\caption{Raw W1 light curves (left column, blue) and W1$-$W2 color trends (right column, green) for six candidate ``cv'' discoveries. See text for details.}
\label{fig:discoveries_cv}
\end{figure}

Figure~\ref{fig:discoveries_cv} shows six candidate ``cv'' discoveries from VarWISE, ordered by RA. These are VarWISE J012736.45+590547.7, VarWISE J013809.18+622826.3, VarWISE J080001.20$-$352210.9, VarWISE J154955.20+332751.8, VarWISE J211206.88 +510346.8, and VarWISE J230627.38+593515.7 in Figure~\ref{fig:discoveries_cv}a-f, respectively. Both the phase-folded W1 light curves and phase-folded W1$-$W2 color trends are shown.

\subsection{Class ``yso''\label{sec:analysis_yso}}

Of the 81 ``yso'' objects selected via the criteria in Table~\ref{tab:analysis_column_selections}, 85\% (69/81) appear to be solid YSO candidates based on their light curve morphologies, red W1$-$W2 colors, and association with nebulosity and/or extinction in the WISE imagery. Another 10\% (8/81) are also likely to be YSOs, but their colors and associations with other embedded stars is less certain. 

A total of 5\% (4/81) appear to be sources with false variability. One of these objects is impacted by a long diffraction spike from a bright star that was not properly flagged in the NEOWISE data; the other three objects are affected by a WISE/NEOWISE pipeline processing issue that is newly identified here. Specifically, when a source falls near a strong gradient in nebulosity, the photometry of that object is sometimes recorded as considerably fainter than reality, and the uncertainties, though larger than normal for that magnitude, fail to capture the extent to which that fainter photometry differs from the truth. Users working on objects in nebulous regions are advised to check their data for this issue.

\begin{figure*}
\includegraphics[width=\linewidth]{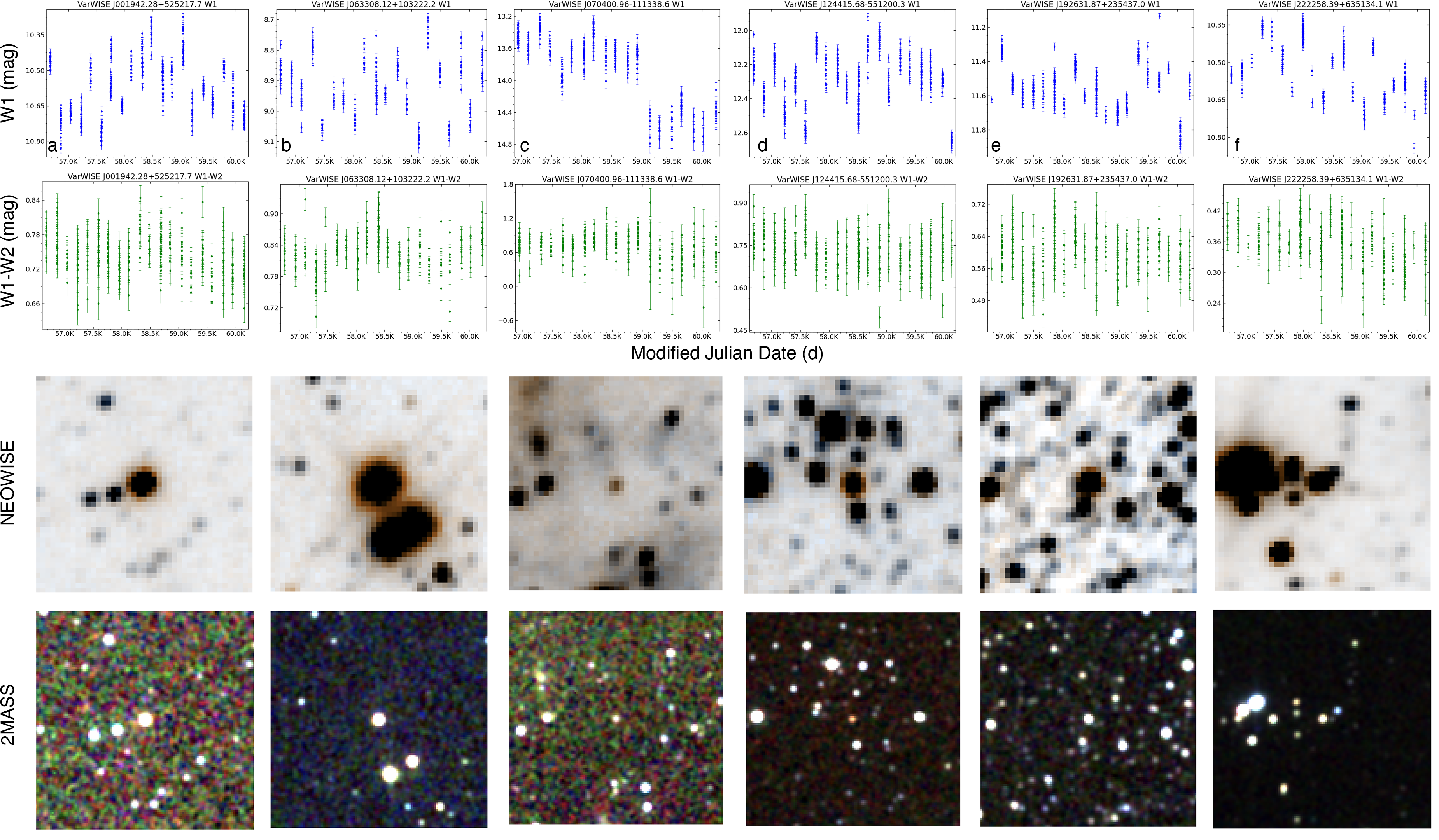}
\caption{Raw W1 light curves, corresponding W1$-$W2 color variations, NEOWISE cutouts, and 2MASS cutouts for six candidate ``yso'' discoveries. See text for details.}
\label{fig:discoveries_yso}
\end{figure*}

Figure~\ref{fig:discoveries_cv} shows six candidate ``yso'' discoveries from VarWISE, ordered by RA. These are VarWISE J001942.28+525217.7, VarWISE J063308.12+103222.2, VarWISE J070400.96$-$111338.6, VarWISE J124415.68$-$551200.3, VarWISE J192631.87 +235437.0, VarWISE J222258.39+635134.1 in Figure~\ref{fig:discoveries_yso}a-f, respectively. The raw W1$-$W2 color variations of these same objects are also shown, as are 2$\times$2-arcmin colors cutouts from NEOWISE (W1+W2) and 2MASS ($J$+$H$+$K_s$). As expected, and as illustrated in the imagery, these objects are much brighter in NEOWISE than in 2MASS and have very red W1$-$W2 colors.

\subsection{Class ``agn''}

Of the 84 ``agn'' objects selected via the criteria in Table~\ref{tab:analysis_column_selections}, 80\% (67/84) have AGN-like light curves and are associated with an object that appears extended (and not in a region of Galactic star formation) in optical or near-infrared surveys such as PanSTARRS\footnote{\url{https://ps1images.stsci.edu/cgi-bin/ps1cutouts}}, the Legacy Surveys\footnote{\url{https://www.legacysurvey.org/viewer}}, or the various surveys by WFCAM\footnote{\url{http://wsa.roe.ac.uk/index.html}}. Another 15\% (13/84) have AGN-like light curves but the optical source, if detected at all, appears to be point-like; these are likely to be unresolved galaxies. Another 4\% (3/84) appear to be Galactic YSOs, and the final 1\% (1/84) appears to be a non-periodic pulsating star in M31.

\begin{figure*}
\includegraphics[width=\linewidth]{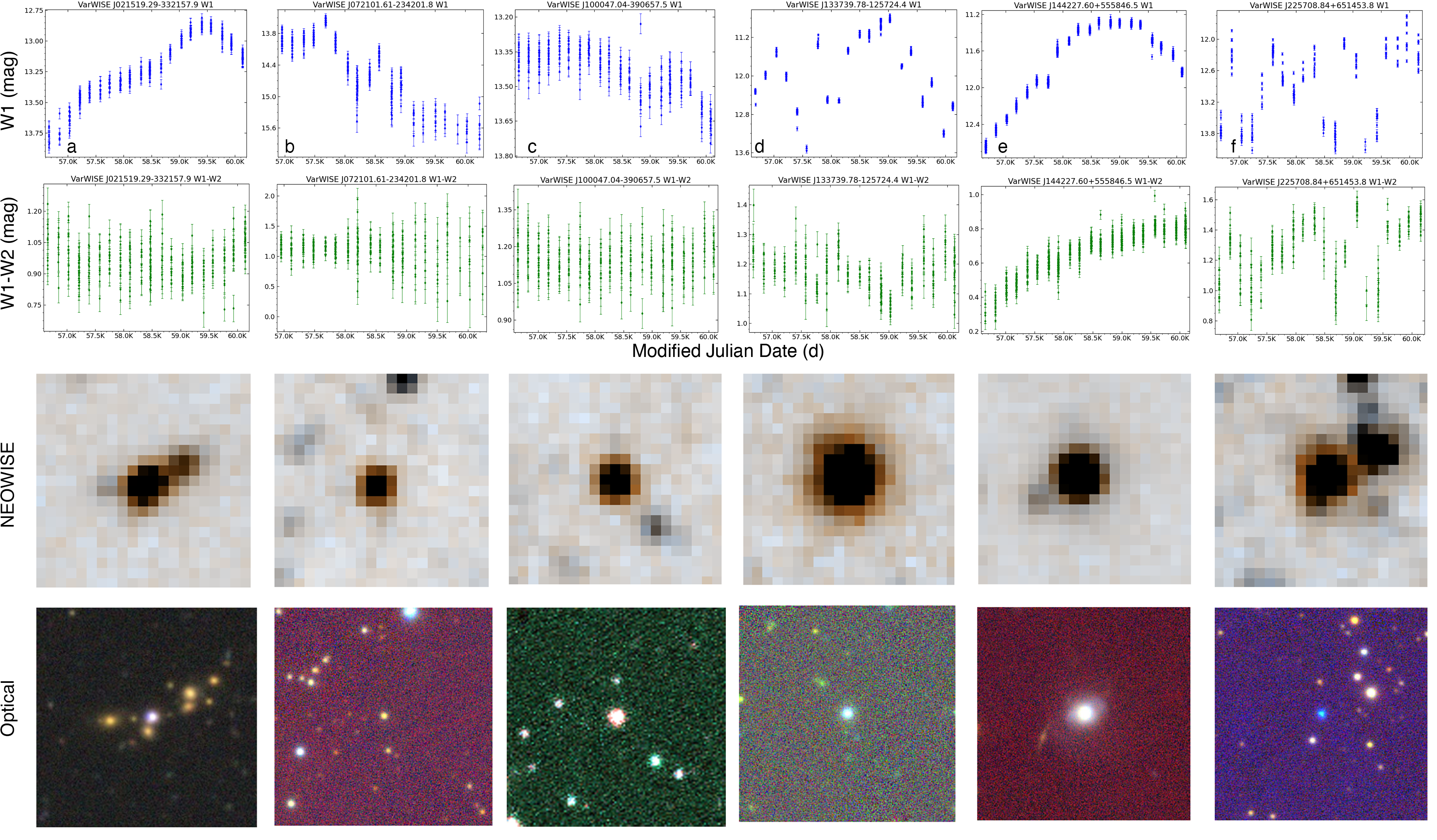}
\caption{Raw W1 light curves, corresponding W1$-$W2 color variations, NEOWISE cutouts, and PanSTARRS/Legacy cutouts for six candidate ``agn'' discoveries. See text for details.}
\label{fig:discoveries_agn}
\end{figure*}

Figure~\ref{fig:discoveries_agn} shows six candidate ``agn'' discoveries from VarWISE, ordered by RA. These are VarWISE J021519.29$-$332157.9, VarWISE J072101.61$-$234201.8, VarWISE J100047.04$-$390657.5, VarWISE J133739.78$-$125724.4, VarWISE J144227.60 +555846.5, and VarWISE J225708.84+651453.8 in Figure~\ref{fig:discoveries_agn}a-f, respectively. The raw W1$-$W2 color variations of these same objects are also shown, as are 1$\times$1-arcmin cutouts from NEOWISE (W1+W2) and either PanSTARRS ($g$+$i$+$y$), the Dark Energy Spectroscopic Experiment (DESI; \citealt{dey2019}) Legacy Imaging Surveys ($g$+$r$+$z$), or SkyMapper (\citealt{onken2024}; $g$+$r$+$z$).

\subsection{Class ``sn''\label{sec:analysis_sn}}

Of the 212 ``sn'' objects selected via the criteria in Table~\ref{tab:analysis_column_selections}, only 9\% (18/212) appear to be solid SN candidates, with another 35\% (75/212) possibly showing slow rises toward or long decays from a supernova event, although most of these are most likely to be normal AGN activity. Most of the remaining 56\% (119/212) appear to be normal AGNs.

\begin{figure*}
\includegraphics[width=\linewidth]{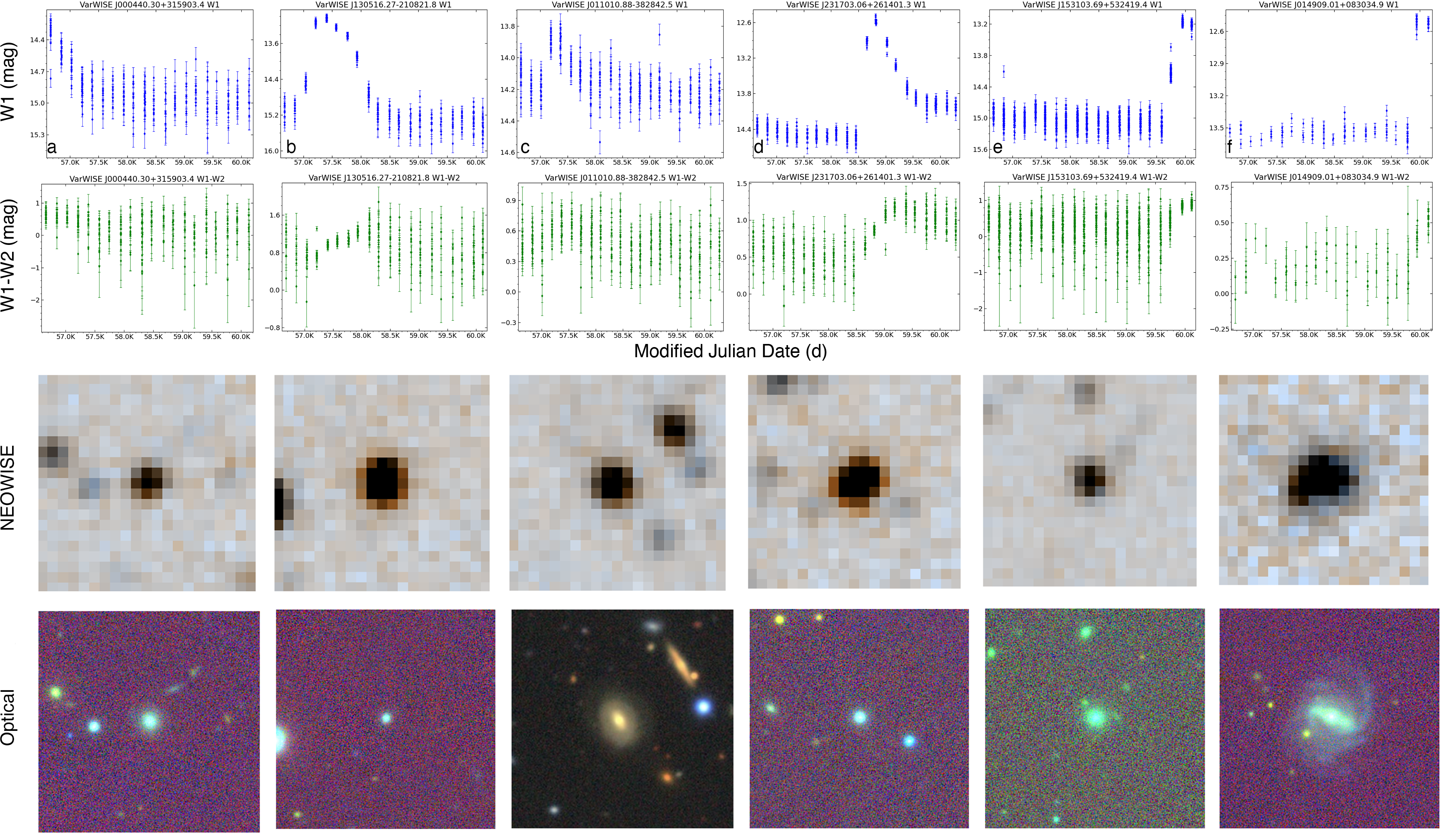}
\caption{Raw W1 light curves, corresponding W1$-$W2 color variations, NEOWISE cutouts, and PanSTARRS/Legacy cutouts for six candidate ``sn'' discoveries. See text for details.}
\label{fig:discoveries_sn}
\end{figure*}

Figure~\ref{fig:discoveries_sn} shows six candidate ``sn'' discoveries from VarWISE, ordered chronologically by the time of the SN-like event. These are VarWISE J000440.30+315903.4, VarWISE J130516.27$-$210821.8, VarWISE J011010.88$-$382842.5, VarWISE J231703.06+261401.3, VarWISE J153103.69 +532419.4, and VarWISE J014909.01+083034.9 in Figure~\ref{fig:discoveries_sn}a-f, respectively. The raw W1$-$W2 color variations of these same objects are also shown, as are 1$\times$1-arcmin cutouts from NEOWISE (W1+W2) and either PanSTARRS ($g$+$i$+$y$) or the DESI Legacy Imaging Surveys ($g$+$r$+$z$).

\section{Conclusions}

The VarWISE time-domain survey provides all-sky coverage from a space-based platform using a decade-long baseline. It provides this capability well into the infrared -- at 3.4 and 4.6 $\mu$m -- meaning that it can uniquely detect objects deeply embedded in dust, both in the plane of the Milky Way and in dust-obscured galaxies outside our own. It also provides a wide dynamic range ($ 7.5 < W1 < 16.0$ mag), enabling variability detection for some objects that are saturated in other time-domain surveys.

The extended version of the catalog contains 1.9M objects, and the purified version of the catalog, containing only those variables of highest confidence, contains 457k objects. These data sets should be a valuable resource for the characterization of known variables and the discovery of new ones -- particularly those most easily seen at these infrared wavelengths. The VarWISE Extended Catalog, VarWISE Pure Catalog, and the Associations Table, which links these variables back to the individual epoch data from the NEOWISE-R Single Exposure (L1b) Source Table, are available at the NASA/IPAC Infrared Science Archive\footnote{\url {https://irsa.ipac.caltech.edu/}}.

\begin{acknowledgements}
    \section{ACKNOWLEDGEMENTS}
    This publication makes use of data products from the Near-Earth Object Wide-field Infrared Survey Explorer (NEOWISE), which is a joint project of the Jet Propulsion Laboratory/California Institute of Technology and the University of Arizona. NEOWISE is funded by the National Aeronautics and Space Administration.

    This publication makes use of data products from the Two Micron All Sky Survey, which is a joint project of the University of Massachusetts and the Infrared Processing and Analysis Center/California Institute of Technology, funded by the National Aeronautics and Space Administration and the National Science Foundation.
\end{acknowledgements}

\bibliography{main}{}
\bibliographystyle{aasjournal}

\appendix
\restartappendixnumbering

\section{Extinction correction and dereddening}

\label{sec:dereddening}

We derive intrinsic colors and absolute magnitudes for each object in Figure~\ref{fig:G_BP-G_RP_vs_M_G_all_classes} by removing the Galactic foreground extinction appropriate to its Gaia distance. Distances are assigned from Gaia astrometry using the geometric distance estimates of \citet{bailer-jones2021} when available; if no geometric estimate is provided, and the fractional parallax uncertainty is sufficiently small, we adopt the inverse parallax. For each object we then determine the cumulative line-of-sight reddening, $E(B{-}V)$, integrated along that sightline out to the object's location, using a prioritized sequence of three-dimensional dust maps.

\begin{itemize}

\item (1) Local ISM ($\lesssim$ few hundred pc):
For nearby stars, we query the three-dimensional tomographic reconstruction of the local dust density by \citet{LeikeEnsslin2020}. This map treats dust as a continuous three-dimensional field and infers the dust density directly from Gaia distances and multi-band stellar colors. The solution resolves structure at $\sim$1\,pc scales and is calibrated out to $\simeq 400$\,pc. It returns the integrated extinction to any specified heliocentric distance. We adopt this solution whenever (i) the source has a Gaia-inferred distance $\leq 400$\,pc and (ii) the \citet{LeikeEnsslin2020} query yields a well-constrained cumulative $E(B{-}V)$ at that distance. This preferentially captures the high-contrast local interstellar medium that dominates extinction for very nearby objects.

\item (2) Galactic plane / kiloparsec regime:
If \citet{LeikeEnsslin2020} is unavailable or poorly constrained at the source distance, we instead evaluate the all-sky three-dimensional Galactic extinction grid of \citealt{Dharmawardena2024} (hereafter \texttt{Dharma2024}). \texttt{Dharma2024} tabulates cumulative $E(B{-}V)$ as a function of Galactic longitude $l$, latitude $b$, and heliocentric distance $d$ on a regular $(l,b,d)$ mesh with $1^\circ \times 1^\circ$ angular sampling and 1.7\,pc sampling along the line of sight. The map extends to a heliocentric distance of 2.8\,kpc and is constructed from an input catalog of 120 million stars with Gaia astrometry and photometry combined with 2MASS and WISE infrared photometry, enabling extinction and distance estimates in heavily reddened regions of the Galactic plane. We evaluate \texttt{Dharma2024} at each source's Gaia distance to obtain the cumulative $E(B{-}V)$ to that location. We prioritize \texttt{Dharma2024} at kiloparsec distances and in low-latitude sightlines because it is explicitly constructed to trace high-extinction inner-disk structure while maintaining uniform 1.7\,pc radial sampling out to 2.8\,kpc.

\item (3) Bayestar fallback:
For stars for which neither \citet{LeikeEnsslin2020} nor \texttt{Dharma2024} yields a reliable cumulative reddening at the Gaia distance, we adopt the Bayestar three-dimensional dust map of \citet{Green2019}. Bayestar uses Gaia parallaxes together with Pan-STARRS1 and 2MASS photometry to infer reddening as a function of distance along each line of sight, and covers most of the sky to a few kiloparsecs. Bayestar is placed after \texttt{Dharma2024} in the priority order because its performance degrades in the most crowded, highest-extinction inner-plane regions and at very small ($\ll 1$\,kpc) distances --- regimes where \texttt{Dharma2024} and \citet{LeikeEnsslin2020} are explicitly tuned.

\item (4) 2D full-column fallback:
If none of the three-dimensional maps returns a usable value at the source distance, we revert to the two-dimensional all-sky reddening map of \citet{SchlaflyFinkbeiner2011}, which reports the total Galactic dust column integrated to infinity along each line of sight. In this final fallback only, we apply a uniform multiplicative factor of 0.86 to the tabulated full-column $E(B{-}V)$, following the recalibration recommended by \citet{SchlaflyFinkbeiner2011} to bring the original SFD normalization onto a consistent scale. For Galactic sources with finite Gaia distances, we then scale that (renormalized) full-column value down to the source distance by assuming an exponential vertical dust layer with scale height $H_{\rm dust} = 150\,\mathrm{pc}$. This treats the Milky Way dust as a thin disk and applies only the fraction of the total column lying in front of the source, rather than assuming that every object sits behind the entire Galactic dust column. The choice $H_{\rm dust} = 150\,\mathrm{pc}$ is consistent with thin-disk dust scale heights of 120--160 pc inferred from Galactic infrared emission and near-infrared extinction mapping \citep[e.g.][]{DrimmelSpergel2001,Marshall2006}.

\end{itemize}

For each source, the adopted $E(B{-}V)$ from this hierarchy (Leike $\rightarrow$ \texttt{Dharma2024} $\rightarrow$ Bayestar $\rightarrow$ scaled \citealt{SchlaflyFinkbeiner2011}) is converted to passband extinctions in the Gaia $G$, $G_{\rm BP}$, and $G_{\rm RP}$ filters, yielding $A_G$, $A_{\rm BP}$, and $A_{\rm RP}$. We assume a Milky Way extinction law with $R_V \simeq 3.1$ \citep{Cardelli1989,Fitzpatrick1999} and apply fixed coefficients to translate $E(B{-}V)$ into $A_\lambda$. Subtracting these extinctions from the observed photometry gives dereddened apparent magnitudes $G_0$, $G_{{\rm BP},0}$, and $G_{{\rm RP},0}$. Combining $G_0$ with the Gaia distance modulus yields the extinction-corrected absolute magnitude $M_G$, and $(G_{\rm BP}-G_{\rm RP})_0$ provides the intrinsic color.

It is important to note that this procedure corrects only for line-of-sight interstellar dust. It does not remove source-local obscuration (e.g.\ disks around young stellar objects, dusty winds in long-period variables, or accretion flows in cataclysmic variables). Those systems may therefore remain internally reddened even after correction.

\end{document}